\documentclass[english,12pt]{article}
\usepackage{amssymb,mathrsfs,epsfig}
\usepackage[cspex,bbgreekl]{mathbbol} 
\usepackage{latexsym}
\let\savedLambda\Lambda
\let\savedGamma\Gamma
\usepackage{amsmath}
\let\Lambda\savedLambda
\let\Gamma\savedGamma
\advance\textheight by 4cm
\advance\topmargin -2cm
\advance\oddsidemargin -1cm
\advance\evensidemargin -1cm
\newcommand{\tr}{\operatorname{tr}}

\newcommand{\A}{{\mathcal{A}}}

\newcommand{\C}{{\mathcal{C}}}

\newcommand{\D}{{\mathcal{D}}}

\newcommand{\F}{{\mathcal{F}}}
\newcommand{\fg}{{\mathfrak{g}}}

\newcommand{\G}{{\mathcal{G}}}
\renewcommand{\H}{{\mathscr{H}}}
\newcommand{\I}{{\mathcal{I}}}
\newcommand{\J}{{\mathcal{J}}}
\newcommand{\K}{{\mathcal{K}}}
\renewcommand{\L}{{\mathcal{L}}}

\newcommand{\M}{{\mathcal{M}}}

\renewcommand{\P}{{\mathscr{P}}}

\newcommand{\R}{{\mathcal{R}}}

\newcommand{\T}{{\mathcal{T}}}

\numberwithin{equation}{section}

\newcommand{\LD}{{\Bbb{L}}}
\newcommand{\KR}{{\stackrel{\circ}{\Bbb{K}}}}

\renewcommand{\H}{{\mathcal{H}}}

\renewcommand{\P}{{\mathcal{P}}}

\renewcommand{\S}{\mathcal{S}}
\newcommand{\oS}{\overset{\;\circ}{\S}{}}

\renewcommand{\S}{{\mathcal{S}}}

\newcommand{\X}{X}

\newcommand{\fd}{{\mathbb{d}}}
\newcommand{\fV}{{\mathbb{V}}}
\newcommand{\fE}{{\mathbb{E}}}

\newcommand{\fGamma}{{\mathbb{\Gamma}}}
\newcommand{\fLambda}{{\mathbb{\Lambda}}}
\newcommand{\fX}{{\mathbb{X}}}
\newcommand{\mysection}[1]{\section{#1} 
\indent \indent}
\newcommand{\mysubsection}[1]{\subsection{#1} 
\indent \indent}

\begin{document}
\title{Topological Quantum Numbers\\ of\\
Relativistic Two-Particle Mixtures}
\author{S. Pru\ss-Hunzinger, S. Rupp, and M. Sorg
\vspace{0.5cm}\\
{\it II.\ Institut f\"ur Theoretische Physik der
Universit\"at Stuttgart,} \\{\it Pfaffenwaldring 57, D-70550 Stuttgart, Germany}\vspace{0.3cm}}
\maketitle
\begin{abstract}
The relativistic two-particle quantum mixtures are studied from the topological point of view. The mixture field variables can be transformed in such a way that a kinematical decoupling of both particle degrees of freedom takes place with a residual coupling of purely algebraic nature (``exchange coupling''). Both separated sets of particle variables induce a certain map of space-time onto the corresponding ``exchange groups'', i.e. $SU(2)$ and $SU(1,1)$, so that for the compact case ($SU(2)$) there arises a pair of winding numbers, either odd or even, which are a topological characteristic of the two-particle Hamiltonian.
\end{abstract}
\newpage
\mysection{Introduction}
Many of the great successes of modern field theory, classical and quantum, are undoubtedly due to the use of topological methods. Indeed, these methods did not only help classifying and subdividing the set of solutions of the field equations, as e.g. in monopole \cite{CrGoNa} and instanton \cite{FeUh,DoKr} theory, but they also generated unexpected new relationships between apparently unrelated branches in particle theory; e.g. it has been observed within the framework of the string theories that the Chern-Simons theory, as a quantum field theory in three dimensions, is intimately related to the invariants of knots and links \cite{Wi,At} which first had been discussed from a purely mathematical point of view \cite{Jo}. The list of examples of a fruitful cooperation of topology and physics could easily be continued but may be sufficient here in order to demonstrate the significance of topological investigations of physical theories.

The present paper is also concerned with such a topological goal, namely to elaborate the topological invariants which are encoded into the Hamiltonian of the ${\Bbb C}^2$ realization of {\it Relativistic Schr\"odinger Theory (RST)}, a recently established alternative approach to relativistic many-particle quantum mechanics \cite{So97,MaRuSo01,RuSo01} (for a critical discussion of some of the deficiencies of the conventional Bethe-Salpeter equation see ref. \cite{Gr}). The present new approach works for an arbitrary number of scalar \cite{MaRuSo01} and spin particles \cite{MaSo99}, but for the present purposes we restrict ourselves to a scalar two-particle system which requires the ${\Bbb C}^2$ realization of RST. This means that the RST wave functions $\Psi(x)$ constitute a section of a complex vector-bundle over space-time as the base space and the two-dimensional complex plane ${\Bbb C}^2$ as the typical fibre. Such a construction clearly displays the main difference to the conventional approach where the Hilbert space of a two-particle system is taken as the tensor product of the one-particle spaces, whereas in RST one takes the Whitney sum of the single-particle bundles. Thus the latter construction suggests a fluid-dynamic interpretation of the formalism in contrast to the conventional approach whose tensor-product construction rather meets with Born's probabilistic interpretation.

Now it should be evident that such differences in the mathematical formalism necessarily will imply also different descriptions of the physical phenomena. For the present topological context, the phenomenon of {\it entanglement} becomes important. In the conventional theory, an entangled $N$-particle system requires the use of either the symmetric or anti-symmetric wave functions $\Psi_{\pm}(x_1,x_2,...x_n)$, see e.g. equation (\ref{e3:3}) below. For the relativisitic treatment, this automatically would introduce $N$ time variables $x^0_1...\,x^0_N$ for such a quantum state $\Psi_{\pm}$ which makes its physical interpretation somewhat obscure (however in the non-relativistic limit, there are no such problems and one may interpret the (anti-)symmetrization postulate simply as a modification of the topology of the configuration space \cite{LeMy}). Naturally, for the Whitney sum construction such a process of (anti-)symmetrization is not feasible because the $N$-particle wave function $\Psi(x)$ remains an object defined at any event $x$ of the underlying space-time. Consequently the conventional matter dichotomy of (anti-)symmetric states must be incorporated into RST in a different way, namely in form of the positive and negative {\it mixtures}. It has already been observed that the physical properties of positive RST mixtures resemble the symmetrized states of the conventional theory whereas the negative mixtures appear as the RST counterpart of the anti-symmetrized conventional states \cite{RuSo01_2}. Thus, in view of such a physical and mathematical dichotomy of the RST solutions, one naturally asks whether perhaps those positive and negative RST mixtures carry different topological features in analogy to the topological modification of the conventional configuration space \cite{LeMy}.

Subsequently these questions are studied in detail and the results are the following: the positive mixtures carry a pair of topological quantum numbers (``winding numbers'') which are either even or odd, whereas the negative mixtures have always trivial winding numbers. This comes about because the positive mixtures induce a map from space-time onto the (compact) group $SU(2)$ whereas the corresponding map of the negative mixtures refers to the (non-compact) group $SU(1,1)$. Clearly the winding number of negative mixtures, as the value of the pullback of the invariant volume form over $SU(1,1)$ upon some three-cycle $C^3$ of space-time, must necessarily produce a trivial result; whereas for the compact case $SU(2)$ one observes the emergence of a pair of even or odd winding numbers.

The procedure to obtain these results is the following: In {\bf Sect. II}, the general theory is briefly sketched for the sake of sufficient self-containment of the paper. {\bf Sect. III} then presents the specialization of the general formalism to the two-particle case to be treated during the remainder of the paper. Here, the emphasis lies on the right parametrization of the theory: it is true, the wave function parametrization reveals the fact that the general two-particle mixture may be described in terms of four single-particle Klein-Gordon wave functions $\{ \psi_1$, $\psi_2$; $\phi_1$, $\phi_2\}$ cf. equations (\ref{e3:70}). These four wave functions undergo ($2+2$)-pairing such that either pair $\{ \psi_1$, $\phi_1\}$ and $\{ \psi_2$, $\phi_2\}$ builds up a four-current $j_{1 \mu}[\psi_1,\phi_1]$ and $j_{2 \mu}[\psi_2,\phi_2]$ as the sources of some vector potential $A_{2 \mu}$ and $A_{1 \mu}$ which then enters the covariant derivatives for the members of the other pair, (\ref{e3:66})-(\ref{e3:67}). However such a wave function parametrization is extremely ineffective for the purpose of detecting the topological peculiarities of the theory, and therefore one resorts to the parametrization in terms of {\it exchange fields}. The point here is that either of the two particle degrees of freedom can be equipped by its own triplet of exchange fields $\{X_{a \mu}$, $\Gamma_{a \mu}$, $\Lambda_{a \mu} \}$ ($a=1,2$) so that the corresponding dynamical equations become decoupled, see the source equations (\ref{e3:36}) and the curl equations (\ref{e3:37}). The entanglement of both particles arises now as an {\it algebraic} constraint upon the two sets of exchange fields, cf. (\ref{e3:40})-(\ref{e3:41}).

The advantage of such a parametrization in terms of exchange fields is made evident in {\bf Sect. IV}, where the curl equations for the exchange fields are identified with the Maurer-Cartan structure equations over the ``exchange groups'', namely the compact $SO(3)$ for the positive mixtures and the non-compact $SO(1,2)$ for the negative mixtures. Consequently the RST exchange fields can be thought to be generated by some map from space-time to the corresponding exchange group so that their curl equations (\ref{e3:37}) are automatically satisfied and it remains to obey the source equations (\ref{e3:36}) together with the exchange coupling conditions (\ref{e3:20})-(\ref{e3:21}). The latter condition can be satisfied by coupling the group elements $\G_a$ for both particles ($a=1,2$) in the right way. For this purpose one has to look for a convenient parametrization of the exchange groups so that the exchange coupling condition adopts a simple form in terms of the selected group parameters. In this context, the optimal parametrization is achieved in terms of the well-known Euler angles $\{ \gamma_1$, $\gamma_2$, $\gamma_3 \}$ relative to which the exchange coupling condition is expressed in an almost trivial way, cf. equation (\ref{e4:28}).

However for the purpose of attaining a concrete geometric picture of the emergence of topological quantum numbers ({\bf Sect. V}), it is more instructive to resort to the universal covering groups, i.e. $SU(2)$ for the positive mixtures (and $SU(1,1)$ for the negative mixtures). The reason is that the group $SU(2)$ is topologically equivalent to the 3-sphere $S^3$ and therefore also to the 3-cycles $C^3$ of space-time. On the other hand, both sets of exchange fields induce a map $\G_{a (x)}$ from space-time into the exchange group, i.e. $SU(2) \sim S^3$ for the positive mixtures, and thereby associate two winding numbers $Z_{(a)}$ ($a=1,2$) (\ref{e5:14}) to any three-cycle $C^3$ as the number of times this cycle is wound onto the exchange group. Since the exchange fields are constituents of the Hamiltonian $\H_{\mu}$, the pair of winding numbers $Z_{(a)}$ are a topological characteristic of the Hamiltonian. As expected, the exchange coupling condition between both particle degrees of freedom should establish some correlation between the winding numbers, but this correlation turns out to be so weak that it simply results in the property of both winding numbers being either odd or even. As an example of a non-trivial exchange field configuration, with unit winding numbers, the method of stereographic projection is applied ({\bf Sect. VI}) which compactifies the Euclidean 3-space $E_3$ (as a time-slice of space-time) to $S^3$ by one-point compactification.

\mysection{General RST Dynamics}
One of the crucial points for the subsequent topological discussions refers to an adequate parametrization of the RST field system. The reason for this is that the topological characteristics of the solutions become manifest only through an adequate choice of the field variables, whereas the general RST dynamics (in operator form) does itself not suggest such a preferred choice. Therefore we briefly list the key features of the general RST dynamics and then concentrate upon a certain parametrization being most helpful for the present topological purposes.

\mysubsection{Matter Dynamics}
The basic building blocks of the theory consist of (i) the matter dynamics, (ii) Hamiltonian dynamics, and (iii) gauge field dynamics \cite{MaRuSo01}. Restricting ourselves here to a system of $N$ scalar particles, one uses an $N$-component wave function $\Psi$ when the particles are ``disentangled'' but one resorts to an Hermitian $N\times N$ matrix $\I$ for the ``entangled'' (i.e. mixture) situation. The motion of matter is then governed by the Relativistic Schr\"odinger Equation (RSE) for the wave function $\Psi$
\begin{equation}
\label{e2:1}
i\hbar c\, \D_{\mu}\Psi=\H_{\mu}\cdot \Psi \; ,
\end{equation}
or by the Relativistic von Neumann Equation (RNE) for the ``{\it intensity matix}'' $\I$
\begin{equation}
\label{e2:2}
\D_{\mu}\I= \frac{i}{\hbar c}\,[\I\cdot\overline{\H}_{\mu}-\H_{\mu}\cdot\I]\;,
\end{equation}
which constitutes the ``{\it matter dynamics}''. The wave functions $\Psi_{(x)}$, to be considered as sections of an appropriate (complex) vector bundle over pseudo-Riemannian space-time as the base space, describe a pure RST state of the N-particle system; whereas the intensity matrix $\I$, as an operator-valued bundle section, is to be used for a mixture configuration. Clearly, the pure states can also be considered as a special kind of mixture, for which the intensity matrix $\I$ then degenerates to a simple tensor product
\begin{equation}
\label{e2:3}
\I \Rightarrow \Psi \otimes  \bar{\Psi} \; .
\end{equation}
Obviously, for such a degenerate situation, the intensity matrix $\I$ must obey the {\it Fierz identity} \cite{MaRuSo01}
\begin{equation}
\label{e2:4}
\I^2-\rho\cdot \I \equiv 0 
\end{equation}
where the scalar density $\rho$ is given by the trace of $\I$
\begin{equation}
\label{e2:5}
\rho \doteqdot \tr \I \; .
\end{equation}

\mysubsection{Hamiltonian Dynamics}
The Hamiltonian $\H_{\mu}$, emerging in the matter dynamics (\ref{e2:1})-(\ref{e2:2}), is itself a (non-Hermitian) dynamical object of the theory and therefore must be determined from its own field equations (``{\it Hamiltonian dynamics}''), namely from the {\it integrability condition}
\begin{equation}
\label{e2:6}
\D_{\mu}\H_{\nu}-\D_{\nu}\H_{\mu}+\frac{i}{\hbar c}[\H_{\mu},\H_{\nu}] = i \hbar c \, \F_{\mu\nu}
\end{equation}
and the {\it conservation equation}
\begin{equation}
\label{e2:7}
\D^{\mu}\H_{\mu}-\frac{i}{\hbar c}\H^{\mu}\cdot\H_{\mu}=-i\hbar c \left(\frac{\M c}{\hbar}\right)^{2}\;.
\end{equation}
Here, the first equation (\ref{e2:6}) guarantees the (local) existence of solutions for the matter dynamics (\ref{e2:1})-(\ref{e2:2}) via the bundle identities for $\Psi$
\begin{equation}
\label{e2:8}
[\D_{\mu}\D_{\nu}-\D_{\nu}\D_{\mu}]\, \Psi=\F_{\mu \nu} \cdot \Psi
\end{equation}
or for $\I$, resp.
\begin{equation}
\label{e2:9}
[\D_{\mu}\D_{\nu}-\D_{\nu}\D_{\mu}]\, \I=[\F_{\mu \nu},\I] \; .
\end{equation}
The meaning of the second equation (\ref{e2:7}) refers to the validity of certain conservation laws, such as charge conservation to be expressed in terms of the current operator $\J_{\mu}$
\begin{equation}
\label{e2:10}
\D^{\mu} \J_{\mu} \equiv 0 \; ,
\end{equation}
or energy-momentum conservation to be written in terms of the energy-momentum operator $\T_{\mu \nu}$ as
\begin{equation}
\label{e2:11}
\D^{\mu} \T_{\mu \nu}+\frac{i}{\hbar c}[\,\overline{\H}^{\mu}\cdot \T_{\mu \nu}-\T_{\mu \nu}\cdot \H^{\mu}]=0 \; .
\end{equation}

\mysubsection{Gauge Field Dynamics}
The third building block of the RST dynamics refers to the gauge field $\F_{\mu \nu}$ (``bundle curvature'') which is defined in terms of the gauge potential $\A_{\mu}$ (``bundle connection'') as usual
\begin{equation}
\label{e2:12}
\F_{\mu \nu} = \nabla_{\mu} \A_{\nu} - \nabla_{\nu} \A_{\mu} + [\A_{\mu},\A_{\nu}] \; .
\end{equation}
The gauge potentials are Lie-algebra valued 1-forms over space-time and do also enter the gauge-covariant derivative $\D$ in the usual way, i.e. for the wave functions $\Psi$ 
\begin{equation}
\label{e2:13}
\D_{\mu} \Psi \doteqdot \partial_{\mu}\Psi+ \A_{\mu} \cdot \Psi \; ,
\end{equation}
or similarly for the operators such as the intensity matrix $\I$
\begin{equation}
\label{e2:14}
\D_{\mu} \I \doteqdot \partial_{\mu}\I+ [\A_{\mu}, \I] 
\end{equation}
or the field strength $\F_{\mu \nu}$ 
\begin{equation}
\label{e2:15}
\D_{\lambda}\F_{\mu \nu} = \nabla_{\lambda} \F_{\mu \nu} + [ \A_{\lambda},\F_{\mu \nu}] \; .
\end{equation}

Now in order to close the RST dynamics for the operators, one has to specify some dynamical equations for the gauge field $\F_{\mu \nu}$. Here, our nearby choice is the (non-abelian generalization of) Maxwell's equation
\begin{equation}
\label{e2:16} 
\D^{\mu} \F_{\mu \nu} = 4 \pi \alpha_{\ast}\, \J_{\nu} \; ,
\end{equation}
\centerline{($\alpha_{\ast}=$ coupling constant).}
This is surely a consistent choice because the generally valid bundle identity 
\begin{equation}
\label{e2:17}
\D^{\mu}\D^{\nu}\F_{\mu \nu} \equiv 0
\end{equation}
immediately implies the charge conservation law (\ref{e2:10}) which is independently implied by the RST dynamics itself. In order to see this in some more detail, one first decomposes the Lie-algebra valued objects whith respect to the (anti-Hermitian) generators $\tau^a$ of the gauge group
\begin{subequations}
\label{e2:18}
\begin{align}
\A_{\mu} &= A_{a \mu} \tau^{a}\\
\F_{\mu \nu} &= F_{a \mu \nu} \tau^{a}\\
\J_{\mu} &= j_{a \mu} \tau^{a} \; .
\end{align}
\end{subequations}
Then one defines the current densities $j_{a \mu}$ in terms of velocity operators $v_{a \mu}$ through
\begin{equation}
\label{e2:19}
j_{a \mu}=\tr (\I\cdot v_{a \mu})
\end{equation}
where the velocity operators read in terms of the Hamiltonian $\H_{\mu}$
\begin{equation}
\label{e2:20}
v_{a \mu} = \frac{i \hbar c}{2}(\, \overline{\H}_{\mu}\cdot (\M c^2)^{-1}\cdot \tau_{a} + \tau_{a}\cdot (\M c^2)^{-1}\cdot \H_{\mu}) \; .
\end{equation}

By this arrangement, the charge conservation law (\ref{e2:10}) reads in component form 
\begin{equation}
\label{e2:21}
D^{\mu}j_{a \mu} \equiv 0 
\end{equation}
with the gauge covariant derivatives being defined in terms of the structure constants $C^{bc}{}_{a}$ of the gauge group through:
\begin{equation}
\label{e2:22}
D_{\mu} j_{a \nu} \doteqdot \nabla_{\mu} j_{a \nu} + C^{bc}{}_{a}\, A_{b
 \mu} j_{c \nu}
\end{equation}
\centerline{($\, [\tau^a,\tau^b]=C^{ab}{}_c \tau^c$).}
However the continuity equations (\ref{e2:21}) are found to be guaranteed just by the matter dynamics (\ref{e2:2}) together with the RST conservation equation (\ref{e2:7}) where the mass operator $\M$ is assumed to be covariantly constant
\begin{equation}
\label{e2:23}
\D_{\mu}\, \M \equiv 0 \; . 
\end{equation}

A similar result does also apply to the energy-momentum conservation (\ref{e2:11}) which in the presence of gauge interactions (as source of energy-momentum) must be generalized to
\begin{equation}
\label{e2:24}
\D^{\mu}\T_{\mu \nu}+\frac{i}{\hbar c}[\,\overline{\H}^{\mu}\cdot \T_{\mu \nu}-\T_{\mu \nu}\cdot \H^{\mu}] = {\mathfrak{f}}_{\nu} \; . 
\end{equation}
Defining here the energy-momentum operator of the scalar RST matter through \cite{MaRuSo01}
\begin{equation}
\label{e2:25}
\T_{\mu \nu} = \frac{1}{2} \left\{ \overline{\H}_{\mu} \cdot (\M c^2)^{-1}\cdot \H_{\nu} +  \overline{\H}_{\nu} \cdot (\M c^2)^{-1}\cdot \H_{\mu} - g_{\mu \nu}[  \overline{\H}^{\lambda} \cdot (\M c^2)^{-1}\cdot \H_{\lambda} - \M c^2] \right\} \; ,
\end{equation}
one finds for the ``{\it force operator}'' ${\mathfrak{f}}_{\nu}$
\begin{equation}
\label{e2:26}
{\mathfrak{f}}_{\nu} = \frac{i \hbar c}{2 }\, [ \overline{\H}^{\mu}\cdot (\M c^2
)^{-1} \cdot \F_{\mu \nu} + \F_{\mu \nu}\cdot (\M c^2)^{-1}\cdot \H^{\mu} ] \; .
\end{equation}
Now one can introduce the ``{\it energy-momentum density}'' $T_{\mu \nu}$ of the RST matter through
\begin{equation}
\label{e2:27}
T_{\mu \nu} \doteqdot \tr \, (\I \cdot \T_{\mu \nu}) 
\end{equation}
the source of which is given by the ``{\it Lorentz force density}'' $f_{\nu}$
\begin{equation}
\label{e2:28}
f_{\nu}  = \tr \ ( \I \cdot {\mathfrak{f}}_{\nu} )=\hbar c F_{a \mu \nu}j^{a \mu}  \; .
\end{equation} 
Clearly if no gauge fields are present ($F_{a \mu \nu} \equiv 0$), the force density $f_{\nu}$ vanishes and the RST matter system is closed
\begin{equation}
\label{e2:29}
\nabla^{\mu} T_{\mu \nu}\equiv 0 \; .
\end{equation}
For non-trivial gauge interactions one can add their energy-momentum density to that of the RST matter and then one also has the closedness relation (\ref{e2:29}) for the total system consisting of RST matter and gauge fields \cite{MaRuSo01}.

In this way, a logically consistent dynamical system for elementary matter has been set up which automatically implies the most significant conservation laws (namely for charge and energy-momentum). However, beyond these pleasant features of the theory, the corresponding solutions carry a non-trivial topology which is not less interesting than their physical implications. This will be demonstrated now for the two-particle systems.

\mysection{Two-Particle Systems}
Once the general RST dynamics has been set up in abstract form, its concrete realizations must be adapted to the spin type and number $N$ of particles as well as to the gauge type of interactions among those particles. The most simple realization is the ${\Bbb{C}}^1$-realization with electromagnetic interactions which is equivalent to the conventional one-particle Klein-Gordon theory \cite{MaSo99_2}. The next complicated case is the ${\Bbb{R}}^2$-realization which may be understood as a kind of embedding theory for the ${\Bbb{C}}^1$-realization so that a point particle becomes equipped with additional structure \cite{RuSiSo00}. Then comes the ${\Bbb{C}}^2$-realization which is adequate for describing systems of two scalar particles, either acted upon by the weak \cite{OcSo96} or electromagnetic interactions \cite{MaRuSo01}. The highest-order realization, which has been considered up to now, is the ${\Bbb{C}}^4$-realization for a Dirac electron subjected to the gravitational interactions \cite{SiSo97}.

\mysubsection{Mixtures and Pure States}
In the present paper we want to study the topological features of scalar two-particle systems with electromagnetic interactions. This means that we have to evoke the ${\Bbb{C}}^2$-realization of RST where the typical vector fibre for the wave functions $\Psi$ is the two-dimensional complex space ${\Bbb{C}}^2$ such that $\Psi$ becomes a two-component wave function
\begin{equation}
\label{e3:1}
\Psi(x)= \left(
\begin{array}{c} \psi_1(x) \\ \psi_2(x) \end{array} 
\right) = \left(
\begin{array}{c} L_1 e^{-i\alpha_1} \\ L_2 e^{-i\alpha_2} \end{array}\; . 
\right)
\end{equation}
Such a field configuration is considered as the RST counterpart of the simple product states of the conventional quantum theory:
\begin{equation}
\label{e3:2}
\Psi_{(1,2)}=\Psi_{(1)}\otimes \Psi_{(2)} \; .
\end{equation}
The entangled states of the conventional theory, either symmetrized or anti-symmetrized
\begin{equation}
\label{e3:3}
\Psi_{\pm (1,2)}=\frac{1}{\sqrt{2}}\big(\Psi_{1(1)}\otimes \Psi_{2(2)}\pm \Psi_{2(1)}\otimes \Psi_{1(2)}\big) \; ,
\end{equation}
have their RST counterparts as the positive and negative mixtures which must be described by an (Hermitian) intensity matrix $\I$ such that $\det \I > 0$ for the positive mixtures and $\det \I < 0$ for the negative mixtures.

Quite analogously as in the conventional theory a superselection rule forbids the transitions between the symmetric and the anti-symmetric states $\Psi_{\pm}$, the RST dynamics also implies such a superselection rule, namely by preserving the sign of $\det \I$ \cite{HuMaSo01}. More concretely, the intensity matrix $\I$ (as any other operator such as $\H_{\mu}$, $\F_{\mu \nu}$, etc.) may be decomposed with respect to a certain operator basis; this then yields the corresponding scalar densities $\rho_a$, $s_a$ ($a=1,2$) as the components of $\I$. Choosing two orthogonal projectors $\P_a$ ($a=1,2$) and two permutators $\Pi_a$ ($a=1,2$) as such an ``orthogonal'' operator basis for the present ${\Bbb{C}}^2$-realization 
\begin{subequations}
\label{e3:4} 
\begin{align}
\P_a \cdot \P_b &= \delta_{ab} \cdot \P_b \\ 
tr \, \P_a &=1 \\
\P_1 + \P_2 &= {\bf 1}\\
\big\{\Pi_a,\P_a \big\} & = \Pi_a \;, \quad \forall a,b \\
\big[ \P_1,\Pi^a \big]&=-\big[\P_2,\Pi^a\big] = i \epsilon^a{}_{b}\, \Pi^b\\
\big\{\Pi_a,\Pi_b\big\}& =2 \delta_{ab} \cdot {\bf 1} \\
\big[ \Pi_a,\Pi_b\big]& =2 i \epsilon_{ab} \, (\P_1-\P_2) \; ,
\end{align}
\end{subequations}
the decomposition of the intensity matrix looks as follows
\begin{equation}
\label{e3:5}
\I=\rho_a \P^a +\frac{1}{2}s_a \, \Pi^a \; .
\end{equation}
\centerline{(summation of indices in juxtaposition).}

Once the scalar densities for the two-particle systems have been defined now, one could convert the general RNE (\ref{e2:2}) into the corresponding field equations for $\rho_a$, $s_a$ (\ref{e3:5}). For an analysis of these field equations it is very instructive to parametrize the four densities $\rho_a$, $s_a$ ($a=1,2$) in terms of three {\it renormalization factors}  $\{Z_T$, $Z_R$, $Z_O\}$ and an amplitude field $L$ 
\begin{subequations}
\label{e3:6} 
\begin{align}
\rho_1+\rho_2\doteqdot \rho &=Z_T \cdot L^2 \\
\rho_1-\rho_2 \doteqdot q &=Z_R \cdot L^2 \\
s=\sqrt{s^as_a} &=Z_O \cdot L^2 \; .
\end{align}
\end{subequations}
Here the amplitude field $L$ emerges through the following observation: when the Hamiltonian $\H_{\mu}$ is decomposed into its Hermitian part $\K_{\mu}$ (``kinetic field'') and anti-Hermitian part $\L_{\mu}$ (``localization field'') according to
\begin{equation}
\label{e3:7}
\H_{\mu}=\hbar c (\K_{\mu}+i\L_{\mu}) \; ,
\end{equation}
then the integrability condition (\ref{e2:6}) says that the trace $L_{\mu}$ of the localization field $\L_{\mu}$ has vanishing curl
\begin{equation}
\label{e3:8}
\nabla_{\mu}L_{\nu}-\nabla_{\nu}L_{\mu}=0 \; .
\end{equation}
\centerline{($L_{\mu}\equiv \tr \L_{\mu}$).}
Therefore $L_{\mu}$ can be generated by some scalar amplitude field $L_{(x)}$ through
\begin{equation}
\label{e3:9}
L_{\mu}=\frac{\partial_{\mu}L^2}{L^2}=2\frac{\partial_{\mu}L}{L}\; ,
\end{equation}
and it is just this amplitude field $L_{(x)}$ (\ref{e3:9}) which has been used for the parametrization (\ref{e3:6}) of the densities $\rho_a$, $s_a$.

With this arrangement, the RNE (\ref{e2:2}) can be transcribed directly to the corresponding field equations for the renormalization factors $\{Z_T$, $Z_R$, $Z_O\}$ which then admit the following first integral
\begin{equation}
\label{e3:10}
Z_T^2-(Z_R^2+Z_O^2)=\sigma_{\ast} \; . 
\end{equation}
Here the ``{\it mixture index}'' $\sigma_{\ast}$ is an integration constant and thus subdivides the density configuration space into two parts: $\sigma_{\ast}=+1$ for the positive mixtures and  $\sigma_{\ast}=-1$ for the negative mixtures; the intermediate value $\sigma_{\ast}=0$ applies to the pure states which mark the border between the two types of mixtures, see fig.1. Observe here that these dynamically disconnected two types of RST mixtures ($\sigma_{\ast}=\pm 1$) are implied by the RST dynamics itself, whereas their conventional analogue (i.e. the interdiction of transitions between the symmetric and anti-symmetric states) is an extra postulate which ``{\it cannot be deduced from the other principles of quantum mechanics}'' \cite{Ba}. In this sense, RST is of less postulative character than the conventional theory.

\mysubsection{Reparametrization}
As we shall readily see, the topological properties of the RST field configurations are rather encoded in the components of the Hamiltonian $\H_{\mu}$, not in the densities $\rho_a$, $s_a$ as the components of the intensity matrix $\I$ (\ref{e3:5}). Here it is important to remark that the Hamiltonian $\H_{\mu}$ decomposes into two conceptually different parts: the single-particle fields and the exchange fields. Naturally the single-particle constituent ${}^{(S)}\H_{\mu}$ of $\H_{\mu}$ is obtained through projection by the single-particle projectors $\P_a$
\begin{equation}
\label{e3:11}
{}^{(S)}\H_{\mu}\doteqdot \P_a \cdot \H_{\mu} \cdot \P^a \; ,
\end{equation}
whereas the exchange part ${}^{(\Pi)}\H_{\mu}$ is the remainder
\begin{equation}
\label{e3:12}
{}^{(\Pi)}\H_{\mu}=\H_{\mu} -{}^{(S)}\H_{\mu} \; .
\end{equation}
Clearly, the entangling exchange interactions between the two particles are contained in the exchange constituent  ${}^{(\Pi)}\H_{\mu}$ (\ref{e3:12}) whereas for a disentangled situation the Hamiltonian $\H_{\mu}$ consists of the single-particle contribution ${}^{(S)}\H_{\mu}$ alone. 

Now, as has been remarked expressly, the right parametrization of the RST field degrees of freedom is crucial for the detection of the topological characteristics. Here it has turned out very helpful to parametrize the exchange part ${}^{(\Pi)}\H_{\mu}$ of the Hamiltonian (\ref{e3:12}) in a redundant way by six exchange fields $\{ X_{a \mu}$, $\Gamma_{a \mu}$, $\Lambda_{a \mu}\}$ ($a=1,2$) obeying the following curl equations \cite{RuSo01_2}
\begin{subequations} 
\label{e3:37}
\begin{align}
\nabla_{\mu}X_{a\nu}-\nabla_{\nu}X_{a\mu}&=2\sigma_{\ast}[\Lambda_{a\mu}\Gamma_{a\nu}-\Lambda_{a\nu}\Gamma_{a\mu}]\\
\nabla_{\mu}\Gamma_{a\nu}-\nabla_{\nu}\Gamma_{a\mu}&= 2\left[X_{a\mu}\Lambda_{a\nu}-X_{a\nu}\Lambda_{a\mu} \right]\\
\nabla_{\mu}\Lambda_{a\nu}-\nabla_{\nu}\Lambda_{a\mu}&=2\left[\Gamma_{a\mu}X_{a\nu}-\Gamma_{a\nu}X_{a\mu} \right] \;.
\end{align}
\end{subequations}
Clearly these curl equations are a direct consequence of the general integrability condition (\ref{e2:6}). The redundancy of this parametrization is expressed by an algebraic constraint upon the exchange fields which effectively reduces their number from six to four, namely for the positive mixtures ($\sigma_{\ast}=+1$)
\begin{subequations}
\label{e3:20}
\begin{align}
X_{1 \mu}&=\sinh \beta \cdot \Gamma_{1 \mu}+\cosh \beta \cdot \Gamma_{2 \mu} \\
X_{2 \mu}&=\sinh \beta \cdot \Gamma_{2 \mu}+\cosh \beta \cdot \Gamma_{1 \mu} \; ,
\end{align}
\end{subequations}
and similarly for the negative mixtures ($\sigma_{\ast}=-1$)
\begin{subequations}
\label{e3:21}
\begin{align}
X_{1 \mu}&=\cosh \beta \cdot \Gamma_{1 \mu}+\sinh \beta \cdot \Gamma_{2 \mu} \\
X_{2 \mu}&=\cosh \beta \cdot \Gamma_{2 \mu}+\sinh \beta \cdot \Gamma_{1 \mu} \; .
\end{align}
\end{subequations}
Here, the mixture variable $\beta$ is merely a transformed version of the former variable $\lambda$ \cite{RuSo01}
\begin{equation}
\label{e3:22}
\lambda \doteqdot \frac{1}{1+\frac{\sigma_{\ast}}{Z_O}}=
\begin{cases}
\frac{1}{2}\Big(1+\tanh \beta\Big),\quad \sigma_{*}=+1\\
\frac{1}{2}\Big(1+\coth \beta\Big),\quad \sigma_{*}=-1\;.
\end{cases} 
\end{equation}

It will readily become evident, in what way the newly introduced exchange triplets $\{X_{a \mu}$, $\Gamma_{a \mu}$, $\Lambda_{a \mu}\}$ lead us directly to the topological properties of the exchange subsystem.

In order to finally close the RST dynamics, we also have to specify the field equation for the mixture variable $\beta$ (\ref{e3:22}), or $\lambda$ resp. In the last end, the desired equation must be traced back to the RNE (\ref{e2:2}) since the mixture variables $\lambda$ (or $\beta$) have been defined in terms of the renormalization factor $Z_O$ (\ref{e3:6}c) which is part of the overlap density $s$ as a component of the intensity matrix $\I$ (\ref{e3:5}). Thus, if one follows the mixture variable through all its reparametrizations, one finally ends up with the following field equation for $\beta$:
\begin{equation}
\label{e3:38}
\partial_{\mu}\beta=
\begin{cases}
2 \cosh{\beta}\left(\Lambda_{1 \mu}+\Lambda_{2 \mu}\right);\quad \sigma_{\ast}=+1\\
2 \sinh{\beta}\left(\Lambda_{1 \mu}+\Lambda_{2 \mu}\right);\quad \sigma_{*}=-1 \; .
\end{cases} 
\end{equation}
Obviously this result identifies the sum of both exchange fields $\Lambda_{1 \mu}+\Lambda_{2 \mu}$  as a gradient field
\begin{subequations}
\label{e3:39}
\begin{align}
\nabla_{\mu}(\Lambda_{1 \nu}+\Lambda_{2 \nu}) -\nabla_{\nu}(\Lambda_{1 \mu}+\Lambda_{2 \mu}) &=0 \\
\Lambda_{1 \mu}+\Lambda_{2 \mu} &=\frac{1}{2}\sigma_{\ast} \partial_{\mu}\epsilon \; .
\end{align}
\end{subequations}
This gradient property could also have been discovered by eliminating the mixture variable $\beta$ from the link between both exchange fields $\X_{a \mu}$ and $\Gamma_{a \mu}$ (\ref{e3:20})-(\ref{e3:21}) which yields two quadratic relations between these fields, namely a symmetric one
\begin{equation}
\label{e3:40}
\Gamma_{1\mu}\Gamma_{1\nu}+\sigma_{*}X_{1\mu}X_{1\nu}=\Gamma_{2\mu}\Gamma_{2\nu}+\sigma_{*}X_{2\mu}X_{2\nu}
\end{equation}
and an anti-symmetric one
\begin{equation}
\label{e3:41}
X_{1\mu} \Gamma_{1\nu}-X_{1\nu}\Gamma_{1\mu}=-\Big[X_{2\mu}\Gamma_{2\nu}-X_{2\nu}\Gamma_{2\mu}\Big]\;.
\end{equation}
However this latter relation just implies the present gradient condition (\ref{e3:39}) when the curl relations (\ref{e3:37}c) for the exchange fields $\Lambda_{a \mu}$ are observed.

For specifying also the source equations for the exchange triplets $\{ X_{a \mu}$, $\Gamma_{a \mu}$, $\Lambda_{a \mu}\}$ one has to observe their coupling to the single-particle subsystem which therefore must first be suitably parametrized. Such a parametrization is obtained by using a doublet of ``{\it kinetic fields}'' $\KR_{a \mu}$ and ``{\it amplitude fields}'' ${\Bbb{L}}_a$ which are required to obey the following ``{\it single particle dynamics}'' \cite{RuSo01_2}

\begin{subequations}
\label{e3:22_kurz}
\begin{align}
\label{e3:35a}
\nabla_{\mu}\KR_{a\nu}-\nabla_{\nu}\KR_{a\mu}&=F_{a\mu\nu} \\
\label{e3:35b}
\nabla^{\mu}(\KR_{a \mu}+X_{a \mu})+2{\Bbb{L}}_a (\KR_{a \mu}+X_{a \mu})&=0\\
\label{e3:32}
\square\LD_{a}+\LD_{a}\Big\{ \Big(\frac{M c}{\hbar}\Big)^{2}-(\KR_a+X_{a \mu})(\KR_a{}^{\mu}+X_a{}^{\mu}) \Big\}&=\sigma_{*}\LD_{a}\big\{\Lambda_{a\mu}{\Lambda_{a}}^{\mu}+\Gamma_{a\mu
}{\Gamma_{a}}^{\mu}\big\}\;.
\end{align}
\end{subequations}
Clearly these source equations are rigorously deducible again from the general conservation equation (\ref{e2:7}) which is the origin also for the two charge conservation laws ($a=1,2$) 
\begin{equation}
\label{e3:30}
\nabla^{\mu}j_{a \mu}=0 \; ,
\end{equation}
following from the general source equations (\ref{e2:21}) for the present case of an abelian gauge group (i.e. $U(1)\times U(1))$. The RST currents $j_{a \mu}$ (\ref{e2:19}) themselves read in terms of the single-particle variables
\begin{equation}
\label{e3:28} 
j_{a \mu} =\frac{\hbar}{Mc}{\Bbb{L}}_a{}^2( \KR_{a \mu} +X_{a \mu})  \; ,
\end{equation}
where the squares of the amplitudes ${\Bbb{L}}_a$ can be identified with the single-particle densities $\rho_a$ (\ref{e3:5}): ${\Bbb{L}}_a{}^2\equiv \rho_a$.

Once an adequate parametrization of the single-particle subsystems has been achieved now, it becomes a straightforward matter to make explicit their coupling to the exchange fields $\{X_{a \mu}$, $\Gamma_{a \mu}$, $\Lambda_{a \mu}\}$:
\begin{subequations}
\label{e3:36}
\begin{align}
\nabla^{\mu}X_{a \mu}&=-2{\Bbb{L}}_a{}^{\mu}X_{a \mu}-\nabla^{\mu}\KR_{a \mu}-2{\Bbb{L}}_a{}^{\mu}\,\KR_{a \mu} \\
\nabla^{\mu}\Gamma_{a\mu}&=-2{\Gamma_{a}}^{\mu}\LD_{a\mu}-2 {\Lambda_{a}}^{\mu}\,
\KR_{a\mu}\\
\nabla^{\mu}\Lambda_{a\mu}&=-2{\Lambda_{a}}^{\mu}\LD_{a\mu}+2{\Gamma_{a}}^{\mu}\,
\KR_{a\mu}\; .
\end{align}
\end{subequations}
These source equations are relevant for treating concrete physical problems (e.g. the Helium problem \cite{RuSo01_2}) but they are not needed for the deduction of the subsequent topological results which are based exclusively upon the curl equations of the exchange fields (\ref{e3:37}).

\mysubsection{Average Exchange Fields}
An alternative way for the redundant description of a system, obeying some constraint in order to remove the redundancy, is to minimalize the number of dynamical variables so that no longer a constraint has to be imposed. Thus, for the present situation one introduces in place of the six exchange fields $\{X_{a \mu}$, $\Gamma_{a \mu}$, $\Lambda_{a \mu}\}$ only three ``{\it average}'' exchange fields $\{ \tilde{\fX}$, $\tilde{\fGamma}$, $\tilde{\fLambda}\}$
\begin{subequations}
\label{e3:26_kurz}
\begin{align}
\tilde{\fX}&=\tilde{X}_{\mu} \fd x^{\mu}\\
\tilde{\fGamma}&=\tilde{\Gamma}_{\mu} \fd x^{\mu}\\
\tilde{\fLambda}&=\tilde{\Lambda}_{\mu} \fd x^{\mu}
\end{align}
\end{subequations}
which are required to obey the curl equations
 \begin{subequations}
\label{e3:59}
\begin{align}
\fd\tilde\fX&=\sigma_{\ast}\, \tilde\fLambda \wedge \tilde \fGamma\\
\fd\tilde\fGamma&=\tilde\fX \wedge \tilde \fLambda \\
\label{e3:61}
\fd\tilde\fLambda&=-4 \tilde\fX \wedge \tilde\fGamma \; .
\end{align}
\end{subequations}
Additionally, one introduces an arbitrary scalar field $\epsilon(x)$ (``{\it exchange angle}'') and if one builds up the original exchange fields $\{\fX_{a}=X_{a \mu} \fd x^{\mu}$, $\fGamma_{a}= \Gamma_{a \mu} \fd x^{\mu}$, $\fLambda_{a}= \Lambda_{a \mu} \fd x^{\mu}\}$ by means of these new variables according to ($\sigma_{\ast}=+1$)
 \begin{subequations}
\label{e3:62}
\begin{align}
\fX_1&=\cos\frac{\epsilon}{2} \cdot \tilde\fX+\sin \frac{\epsilon}{2} \cdot \tilde\fGamma\\
\fGamma_1&=\cos\frac{\epsilon}{2} \cdot \tilde\fGamma-\sin \frac{\epsilon}{2} \cdot \tilde\fX\\
\fLambda_1 &= \frac{1}{2}\, \{ \tilde\fLambda+\frac{1}{2}\, \fd\epsilon \}\\
\fX_2&=\cos\frac{\epsilon}{2} \cdot \tilde\fGamma+\sin \frac{\epsilon}{2} \cdot \tilde\fX\\
\fGamma_2&=\cos\frac{\epsilon}{2} \cdot \tilde\fX-\sin \frac{\epsilon}{2} \cdot \tilde\fGamma\\
\fLambda_2 &= -\frac{1}{2}\, \{ \tilde\fLambda-\frac{1}{2}\, \fd\epsilon \} \; ,
\end{align}
\end{subequations}
then {\it both} the curl equations (\ref{e3:37})
 \begin{subequations}
\label{e3:29_kurz}
\begin{align}
\fd \fX_a &=2\sigma_{\ast} \fLambda_a \wedge \fGamma_a\\
\fd \fGamma_a&=2 \fX_a \wedge \fLambda_a\\
\fd \fLambda_a&=2 \fGamma_a\wedge \fX_a
\end{align}
\end{subequations}
{\it and} the exchange constraints (\ref{e3:40})-(\ref{e3:41}) are satisfied simultaneously! For the negative mixtures ($\sigma_{\ast}=-1$) a similar parametrization in terms of the average fields and exchange angle $\epsilon$ is possible but this is not presented here on account of the topological trivialtity of the negative mixtures.

\mysubsection{Wave Function Description}
Up to now, the RST field system has been parametrized by the Hamiltonian component fields and the densities as the components of the intensity matrix $\I$. Such a parametrization is very helpful for deducing the subsequent topological results, however for practical purposes in view of the physical applications of the theory it is more convenient to parametrize the RST system by means of wave functions. Moreover the latter parametrization yields further insight into the relationship of the new theory with the conventional quantum theory which per se is based upon the concept of wave functions. 

In place of the densities $\rho_a$, $s_a$ as the components of the intensity matrix $\I$ (\ref{e3:5}), the latter operator can also be parametrized in the general case by two two-particle wave functions $\Psi$ and $\Phi$ of the kind (\ref{e3:1}):
\begin{equation}
\label{e3:64}
\I=I_{11} \Psi \otimes \overline{\Psi} +I_{12} \Psi \otimes \overline{\Phi}+I_{21} \Phi \otimes \overline{\Psi}+I_{22} \Phi \otimes \overline{\Phi} \; .
\end{equation}
Here the constants $I_{ab}$ ($a,b=1,2$) must form a Hermitian matrix $\I_{\ast}$ (i.e. $\overline{\I}_{\ast}=\I_{\ast}$) in order that the intensity matrix $\I$ be also Hermitian. The RNE (\ref{e2:2}) for the intensity matrix $\I$ is obeyed whenever either one of the two wave functions $\Psi$ and $\Phi$ obeys the RSE (\ref{e2:1})
 \begin{subequations}
\label{e3:65}
\begin{align}
i\hbar c \D_{\mu}\Psi&=\H_{\mu}\Psi\\
i \hbar c \D_{\mu}\Phi&=\H_{\mu}\Phi \; .
\end{align}
\end{subequations}
Here the covariant derivative (\ref{e2:13}) of the two-particle wave functions reads in components
\begin{equation}
\label{e3:66}
\D_{\mu}\Psi = \left(
\begin{array}{c}
D_{\mu} \psi_1\\ D_{\mu} \psi_2 
\end{array} \right)
= \left(
\begin{array}{c}
\partial_{\mu}\psi_1-iA_{1 \mu}\psi_1 \\
\partial_{\mu}\psi_2-iA_{2 \mu}\psi_2
\end{array}
\right) \; ,
\end{equation}
resp. for the second wave function $\Phi$
\begin{equation}
\label{e3:67}
\D_{\mu}\Phi = \left(
\begin{array}{c}
D_{\mu} \phi_1\\ D_{\mu} \phi_2 
\end{array} \right)
= \left(
\begin{array}{c}
\partial_{\mu}\phi_1-iA_{1 \mu}\phi_1 \\
\partial_{\mu}\phi_2-iA_{2 \mu}\phi_2
\end{array}
\right) \; ,
\end{equation}
where the gauge potential $\A_{\mu}$ (\ref{e2:18}a) has been decomposed with respect to the two $U(1) \times U(1)$ generators $\tau_a$
\begin{equation}
\label{e3:68}
\tau_a=-i\P_a \; .
\end{equation}
Observe here that the first components $\psi_1$, $\phi_1$ are acted upon by the first gauge potential $A_{1 \mu}$ whereas the second potential $A_{2 \mu}$ acts upon the second components $\psi_2$, $\phi_2$.

Now one can easily show by differentiating once more the RSE's (\ref{e3:65}) and using the conservation equation (\ref{e2:7}) for the Hamiltonian $\H_{\mu}$ that the two-particle wave functions $\Psi$ and $\Phi$ must obey the Klein-Gordon equation (KGE)
\begin{subequations}
\label{e3:69}
\begin{align}
\D^{\mu}\D_{\mu}\Psi+\Big(\frac{\M c}{\hbar}\Big)^2 \Psi&=0\\
\D^{\mu}\D_{\mu}\Phi+\Big(\frac{\M c}{\hbar}\Big)^2 \Phi&=0 \; .
\end{align}
\end{subequations}
In components, this result says that the general intensity matrix $\I$ is composed of four single-particle wave functions $\psi_1$, $\psi_2$, $\phi_1$, $\phi_2$ which obey the Klein-Gordon equations
\begin{subequations}
\label{e3:70}
\begin{align}
D^{\mu}D_{\mu}\psi_1+\Big(\frac{M_1 c}{\hbar}\Big)^2 \psi_1&=0\\
D^{\mu}D_{\mu}\phi_1+\Big(\frac{M_1 c}{\hbar}\Big)^2 \phi_1&=0 \\
D^{\mu}D_{\mu}\psi_2+\Big(\frac{M_2 c}{\hbar}\Big)^2 \psi_2&=0\\
D^{\mu}D_{\mu}\phi_2+\Big(\frac{M_2 c}{\hbar}\Big)^2 \phi_2&=0\; .
\end{align}
\end{subequations}
Observe here that the first two equations (\ref{e3:70}a)-(\ref{e3:70}b) are governed by the gauge potential $A_{1 \mu}$ whereas the second half (\ref{e3:70}c)-(\ref{e3:70}d) is referred to the gauge potential $A_{2 \mu}$. Thus the interactions in our two-particle mixture become fixed by specifying the way in which the potentials $A_{a \mu}$ ($a=1,2$) are tight up to the RST currents $j_{a \mu}$, see below.

It has been demonstrated that, in order to avoid unphysical self-interactions \cite{MaRuSo01}, the link between the potentials and currents must be made in such a way that the curvature $F_{1 \mu \nu}$ of the first particle couples to the second current $j_{2 \mu}$
\begin{equation}
\label{e3:71}
\nabla^{\mu}F_{1 \mu \nu}=4 \pi \alpha_{\ast} j_{2 \nu}
\end{equation}
and vice versa for the second curvature component
\begin{equation}
\label{e3:72}
\nabla^{\mu}F_{2 \mu \nu}=4 \pi \alpha_{\ast} j_{1 \nu} \; .
\end{equation}
Thus it remains to be shown in what way both currents $j_{a \mu}$ ($a=1,2$) of the two-particle mixture are built up by the single-particle wave functions $\psi_a$, $\phi_a$. However this connection between wave functions and currents has already been clarified by the former prescriptions (\ref{e2:19})-(\ref{e2:20}); one simply has to insert there the present form of the intensity matrix $\I$ (\ref{e3:64}) in order to find
\begin{eqnarray}
\label{e3:73}
j_{a \mu}=i\frac{\hbar}{2Mc}\big\{ &I_{11}&[\psi^{\ast}_a(D_{\mu}\psi_a)-(D_{\mu}\psi^{\ast}_a)\psi_a]\nonumber \\
+ &I_{12}&[\phi^{\ast}_a(D_{\mu}\psi_a)-(D_{\mu}\phi^{\ast}_a)\psi_a]\nonumber \\
+ &I_{21}&[\psi^{\ast}_a(D_{\mu}\phi_a)-(D_{\mu}\psi^{\ast}_a)\phi_a]\nonumber \\ 
+&I_{22}&[\phi^{\ast}_a(D_{\mu}\phi_a)-(D_{\mu}\phi^{\ast}_a)\phi_a]\big\} \; .
\end{eqnarray}
By prescribing special values to the constant matrix elements $I_{ab}$ (\ref{e3:64}) one can show that the conventional Hartree and Hartree-fock approaches just form the non-relativistic approximations of the present two-particle RST (see a separate paper). Obviously in this wave-function picture, the RST entanglement consists in the fact that the sources $j_{a \mu}$ of the field strengths $F_{a \mu \nu}$ are a mixture of one-particle and interference currents.

It is instructive also to observe the degeneration of the general intensity matrix $\I$ to its pure-state form (\ref{e2:3}) 
\begin{equation}
\label{e3:74}
\I \Rightarrow \Psi^{\prime}\otimes \overline{\Psi}^{\prime}
\end{equation}
which is the tensor product of a two-particle state $\Psi^{\prime}$ and its Hermitian conjugate $\overline{\Psi}^{\prime}$. Such a degeneration occurs when the constant matrix $\I_{\ast}=\{I_{ab}\}$ (\ref{e3:64}) becomes itself degenerate, i.e
\begin{equation}
\label{e3:75}
\det \I_{\ast} \equiv I_{11}I_{22}-I_{12}I_{21}=0 \; .
\end{equation}
For such a situation one can parametrize the four constant matrix elements $I_{ab}$ by two complex numbers $p$ and $b$ such that
\begin{equation}
\label{e3:76}
\begin{array}{lr}
I_{11}=p^{\ast}p \quad&\quad I_{12}=b^{\ast}p \\
I_{21}=p^{\ast}b \quad&\quad I_{22}=b^{\ast}b
\end{array} \; .
\end{equation} 
This arrangement gives rise to recollect the four Klein-Gordon states $\{\psi_1$, $\psi_2$; $\phi_1$, $\phi_2\}$ into only two pure one-particle states $\psi_1^{\prime}$, $\psi_2^{\prime}$ according to
\begin{subequations}
\label{e3:77}
\begin{align}
\psi_1^{\prime}&=p\psi_1+b\phi_1\\
\psi_2^{\prime}&=p\psi_2+b\phi_2 \; .
\end{align}
\end{subequations}
The meaning of the new states (\ref{e3:77}) is immediately evident by introducing the restricted form of the matrix $\I_{\ast}$ (\ref{e3:76}) into the RST currents (\ref{e3:73}) which yields
\begin{subequations}
\label{e3:78}
\begin{align}
j_{1 \mu}& \Rightarrow j_{1 \mu}^{\prime}=i\frac{\hbar}{2Mc} \big[\psi_1^{\ast \prime} (D_{\mu} \psi_1^{\prime})-(D_{\mu}\psi_1^{\ast \prime})\psi_1^{\prime} \big]\\
j_{2 \mu}& \Rightarrow j_{2 \mu}^{\prime}=i\frac{\hbar}{2Mc} \big[\psi_2^{\ast \prime} (D_{\mu} \psi_2^{\prime})-(D_{\mu}\psi_2^{\ast \prime})\psi_2^{\prime} \big] \; .
\end{align}
\end{subequations}

\mysection{Maurer-Cartan Forms}
The curl relations (\ref{e3:29_kurz}) for the single-particle exchange fields $\{\fX_{a}$, $\fGamma_{a}$, $\fLambda_{a}\}$, as well as the curl equations for the average exchange fields $\{\tilde\fX$, $\tilde\fGamma$, $\tilde\fLambda\}$ (\ref{e3:59}), are of a certain well-known structure in mathematics which is known as {\it Maurer-Cartan structure equations} \cite{GoSc}. The exploitation of this mathematical structure yields a deeper insight into the RST subsystem of exchange fields. The Maurer-Cartan forms ($\T^j$, say; $j=1..$dim$\,\fg$) are left-invariant 1-forms over a Lie group $G$ with Lie algebra $\fg$. Let $\fg$ be spanned by a set $\{ \tau_k \}$ of left-invariant vector fields ($k=1..$dim$\, \fg$) with structure constants $C^l{}_{jk}$ 
\begin{equation}
\label{e4:1}
[\tau_j,\tau_k]=C^l{}_{jk}\tau_l \; .
\end{equation}
The Maurer-Cartan forms $\T^j$ may then be taken as the dual objects of the Lie algebra generators $\tau_k$, i.e. the values of $\T^j$ upon $\tau_k$ is then given as usual by
\begin{equation}
\label{e4:2}
<\T^j|\tau_k>=\delta^j{}_k \; .
\end{equation}

It is possible to equip the Lie algebra $\fg$ (as a linear vector space) with a metric $g$, the {\it Killing-Cartan form}, such that any pair $\tau_j$, $\tau_k$ of generators is mapped into a real (or complex) number $g_{jk}$
\begin{equation}
\label{e4:3}
g(\tau_j,\tau_k)=g_{jk} \; .
\end{equation}
Such a metric can be realized by means of the adjoint representation of the generators, i.e. one puts
\begin{equation}
\label{e4:4}
g_{jk}=-\frac{1}{\mbox{dim}\, {\mathfrak{g}}}\cdot \tr \{({\mathfrak{Ad}}\tau_j)\cdot({\mathfrak{Ad}}\tau_k)\}
\end{equation}
where the adjoint representation is given by
\begin{equation}
\label{e4:5}
({\mathfrak{Ad}}\tau_j)^k{}_l=C^k{}_{jl} \; .
\end{equation}
Alternatively one can look upon the metric $g$ as a bijective map from the Lie algebra $\mathfrak{g}$ to its dual space $\overline{\mathfrak{g}}$, in which the Maurer-Cartan forms are living
\begin{eqnarray}
\label{e4:6}
g: \quad \mathfrak{g} \Rightarrow \overline{\mathfrak{g}} \nonumber \\
g^{-1}: \quad  \overline{\mathfrak{g}}\Rightarrow\mathfrak{g}
\end{eqnarray}
i.e.
\begin{eqnarray}
\label{e4:7}
g(\tau_j)=g_{jk}\T^k \\
g^{-1}(\T^j)=g^{jk}\tau_k \nonumber\\
(g^{jk}g_{kl}=\delta^j{}_l)\; . \nonumber
\end{eqnarray}
Consequently, one can realize the duality relations (\ref{e4:2}) by means of the Killing-Cartan form (\ref{e4:4}), namely by putting
\begin{equation}
\label{e4:8}
<\T^j|\tau_k>=g(g^{-1}(\T^j), \tau_k)=g^{jl}g(\tau_l,\tau_k)=g^{jl}g_{lk}=\delta^j{}_k \; .
\end{equation}

In the present context, the point with the Maurer-Cartan forms is now that they obey the structure equations \cite{GoSc}
\begin{equation}
\label{e4:9}
\fd\T^j=-\frac{1}{2}\, C^j{}_{kl}\, \T^k \wedge \T^l \; .
\end{equation}
This looks already very similar to the curl relations for the exchange fields $\tilde\fX_{a}$, $\tilde\fGamma_{a}$, $\tilde\fLambda_{a}$ (\ref{e3:59}) or also for the fields $\fX_{a}$, $\fGamma_{a}$, $\fLambda_{a}$ (\ref{e3:29_kurz}). Thus one may suppose that one could generate the desired exchange fields via the Maurer-Cartan forms over suitable Lie groups $G$. This method would then consist in establishing a suitable map from space-time onto the group $G$ ($x \Rightarrow \G_{(x)} \in G$) and considering the corresponding pullback of the Maurer-Cartan forms from $G$ to space-time. More concretely, let the group element $\G$ corresponding to the space-time event $x$ be $\G_{(x)}$, the Maurer-Cartan pullback $\tilde{\C}$ appears then in terms of $\G_{(x)}$ as
\begin{equation}
\label{e4:10}
\tilde{\C}=\G\cdot \fd\G^{-1}=\C_{\mu}\fd x^{\mu} \; ,
\end{equation}
i.e. in components (exploiting the isomorphism of $\fg$ and its dual $\overline{\fg}$ via the metric map (\ref{e4:6})
\begin{equation}
\label{e4:11}
\C_{\mu}=E^j{}_{\mu}\tau_j=\G_{(x)}\cdot \partial_{\mu} \G_{(x)}{}^{-1} \; .
\end{equation}
On behalf of the structure equations (\ref{e4:9}) the Maurer-Cartan pullback $\tilde{\C}=\fE^j\tau_j$ obeys the relation
\begin{equation}
\label{e4:12}
\fd\tilde{\C}+\tilde{\C}\wedge\tilde{\C}=0 \; ,
\end{equation}
or written in the component fields $\fE^j=\fE^j{}_{\mu}\fd x^{\mu}$
\begin{equation}
\label{e4:13}
\fd \fE^j=-\frac{1}{2} C^j{}_{kl}\fE^k\wedge\fE^l \; .
\end{equation}

Thus we are left with the problem of finding the right groups $G$ in order to identify their Maurer-Cartan pullbacks $\fE^j$ with our former exchange fields, with the corresponding curl equations being then satisfied automatically.

\mysubsection{Exchange Groups}
It should not come as a surprise that the ordinary rotation group $SO(3)$ is the appropriate group for the positive mixtures and similarly the Lorentz group $SO(1,2)$ in ($1+2$) dimensions is adequate for the negative mixtures. The reason for this is that the structure constants for both groups can be taken as the totally anti-symmetric permutation tensor $\epsilon_{ijk}$ ($\epsilon_{123}=+1$), cf. (\ref{e4:1})
\begin{equation}
\label{e4:14}
[\tau_j,\tau_k]=\epsilon^l{}_{jk}\tau_l
\end{equation}
such that the components of the generators become for their adjoint representation (\ref{e4:5})
\begin{equation}
\label{e4:15}
(\tau_j)^k{}_l=\epsilon^k{}_{jl} \; .
\end{equation}
The corresponding Killing-Cartan metric $g_{jk}$ (\ref{e4:4}) turns out as the (pseudo-)Euclidean case $\eta_{jk}$
\begin{equation}
\label{e4:16}
\{ \eta_{jk}\}=
\left(
\begin{array}{ccc} 1&0&0 \\ 0&\sigma_{\ast}&0\\0&0& \sigma_{\ast}
\end{array} 
\right)
\end{equation}
\centerline{($\epsilon^k{}_{jl}=\eta^{ki}\epsilon_{ijl}$, etc.).}

By this arrangement, the comparison of the former curl equations for the exchange fields $\fX_a$, $\fGamma_a$, $\fLambda_a$ (\ref{e3:29_kurz}) with the present Maurer-Cartan structure equations (\ref{e4:13}) admit the following two identifications
\begin{subequations}
\label{e4:17}
\begin{align}
\fE_{(a)}{}^1 &=-2\sigma_{\ast}\fX_a\\
\fE_{(a)}{}^2 &=\mp 2 \fGamma_a\\
\fE_{(a)}{}^3 &=\pm 2\fLambda_a \; .
\end{align}
\end{subequations}
Furthermore, one can also compare the structure equations (\ref{e4:13}) to the former curl equations for the average exchange fields $\tilde{\fX}$, $\tilde\fGamma$, $\tilde\fLambda$ (\ref{e3:59}) which again allows us to identify these fields with certain Maurer-Cartan forms $\tilde\fE^j$ due to both groups $SO(3)$ and $SO(1,2)$
\begin{subequations}
\label{e4:18}
\begin{align}
\tilde\fE^1 &\equiv -2\sigma_{\ast}\tilde\fX\\
\tilde\fE^2 &\equiv 2 \tilde\fGamma\\
\tilde\fE^3 &\equiv 2 \tilde\fLambda \; .
\end{align}
\end{subequations}

\mysubsection{Euler Angles}
For the subsequent topological discussion a very helpful parametrization of the Maurer-Cartan forms (and thus also of the exchange fields) is obtained by specifying the generating group element $\G_{(x)}$ (\ref{e4:10}) in terms of the ``{\it Euler angles}'' $\gamma_j$ ($j,k,l \ =1,2,3$) \cite{ChDeDi}
\begin{equation}
\label{e4:19}
\G_{(\gamma_1,\gamma_2,\gamma_3)}=\G_{1(\gamma_1)}\cdot \G_{2(\gamma_2)}\cdot \G_{3(\gamma_3)} \; .
\end{equation}
Here the individual factors $\G_{j(\gamma_j)}$ are chosen as follows:
\begin{subequations}
\label{e4:20}
\begin{align}
\G_{1(\gamma_1)}&=\exp[\gamma_1\cdot \tau_3]\\
\G_{2(\gamma_2)}&=\exp[\gamma_2\cdot \tau_2]\\
\G_{3(\gamma_3)}&=\exp[\gamma_3\cdot \tau_3] \; ,
\end{align}
\end{subequations}
and then the components $\fE^j$ (\ref{e4:13}) of the Maurer-Cartan form $\tilde{\C}$ (\ref{e4:10}) are immediately obtained in terms of the Euler angles $\gamma_j$, namely for the positive mixtures as follows ($\sigma_{\ast}=+1$)
\begin{subequations}
\label{e4:21}
\begin{align}
\fE^1 &=\sin \gamma_1 \cdot \fd \gamma_2-\cos \gamma_1\cdot \sin \gamma_2\cdot \fd \gamma_3\\
\fE^2 &=-\sin \gamma_1 \cdot \sin \gamma_2\cdot \fd \gamma_3 - \cos \gamma_1\cdot \fd\gamma_2\\
\fE^3 &=-\fd\gamma_1-\cos \gamma_2\cdot \fd\gamma_3\; ,
\end{align}
\end{subequations}
and similarly for the negative mixtures ($\sigma_{\ast}=-1$)
\begin{subequations}
\label{e4:22}
\begin{align}
\fE^1 &=\sinh \gamma_1 \cdot \fd \gamma_2-\cosh \gamma_1\cdot \sinh \gamma_2\cdot \fd \gamma_3\\
\fE^2 &=\sinh \gamma_1 \cdot \sinh \gamma_2\cdot \fd \gamma_3 - \cosh \gamma_1\cdot \fd\gamma_2\\
\fE^3 &=-\fd\gamma_1-\cosh \gamma_2\cdot \fd\gamma_3\; .
\end{align}
\end{subequations}

Observe here that for the negative mixtures ($\sigma_{\ast}=-1$) there exists a second inequivalent parametrization by Euler angles in addition to the first parametrization (\ref{e4:20}), namely
\begin{subequations}
\label{e4:23}
\begin{align}
\G_{1(\gamma_1)}&=\exp[\gamma_1\cdot \tau_3]\\
\G_{2(\gamma_2)}&=\exp[\gamma_2\cdot \tau_1]\\
\G_{3(\gamma_3)}&=\exp[\gamma_3\cdot \tau_3] \; .
\end{align}
\end{subequations}
Due to this latter parametrization the Maurer-Cartan forms for the negative mixtures adopt a somewhat different shape, i.e.
\begin{subequations}
\label{e4:24}
\begin{align}
\fE^1 &=-\{\cosh \gamma_1 \cdot \fd \gamma_2-\sinh \gamma_1\cdot \sin \gamma_2\cdot \fd \gamma_3\}\\
\fE^2 &=\sinh \gamma_1\cdot \fd \gamma_2 - \cosh  \gamma_1 \cdot \sin \gamma_2\cdot \fd\gamma_3\\
\fE^3 &=-\fd\gamma_1-\cos \gamma_2\cdot \fd\gamma_3\; .
\end{align}
\end{subequations}
The discussion of the negative mixtures runs quite analogously as for the positive mixtures, but since the negative-mixture topology is trivial we restrict ourselves to the positive case exclusively.

The point with the introduction of the Maurer-Cartan forms is now that our exchange fields $\fX_a$, $\fGamma_a$, $\fLambda_a$ can be expressed in terms of the Euler angles $\gamma_{j(a)}$ for either particle ($a=1,2$) and thus the exchange coupling conditions (\ref{e3:20})-(\ref{e3:21}) can be transcribed to the corresponding coupling conditions for the Euler angles $\gamma_{j(1)}$ (first particle) and $\gamma_{j(2)}$ (second particle). This however gives a very simple result as we shall see readily. For the positive mixtures ($\sigma_{\ast}=+1$) the average exchange fields $\tilde\fX$, $\tilde\fGamma$, $\tilde\fLambda$ adopt the following forms by simply combining the Maurer-Cartan identifications (\ref{e4:18}) with the Euler angle parametrization (\ref{e4:21})
\begin{subequations}
\label{e4:25}
\begin{align}
\tilde\fX&\equiv-\frac{1}{2}\,\tilde\fE^1 =-\frac{1}{2}\,\{\sin \tilde{\gamma_1} \cdot \fd \tilde{\gamma_2}-\cos \tilde{\gamma_1}\cdot \sin \tilde{\gamma_2}\cdot \fd \tilde{\gamma_3}\}\\
\tilde\fGamma&\equiv -\frac{1}{2}\,\tilde\fE^2 = \frac{1}{2}\,\{\sin \tilde{\gamma_1} \cdot \sin \tilde{\gamma_2}\cdot \fd \tilde{\gamma_3} + \cos \tilde{\gamma_1}\cdot \fd\tilde{\gamma_2}\}\\
\tilde\fLambda&\equiv\tilde\fE^3 =-\fd\tilde{\gamma_1}-\cos \tilde{\gamma_2}\cdot \fd\tilde{\gamma_3}\; .
\end{align}
\end{subequations}
By use of this result, the single-particle exchange fields $\fX_a$, $\fGamma_a$, $\fLambda_a$ (\ref{e3:62}) look then as follows
\begin{subequations}
\label{e4:26}
\begin{align}
\fX_1&=-\frac{1}{2}\,\{\sin (\tilde{\gamma_1}-\frac{\epsilon}{2}) \cdot \fd \tilde{\gamma_2}- \sin \tilde{\gamma_2}\cdot \cos( \tilde{\gamma_1}-\frac{\epsilon}{2})\cdot \fd \tilde{\gamma_3}\}\\
\fGamma_1&=\frac{1}{2}\,\{\cos (\tilde{\gamma_1}-\frac{\epsilon}{2})\cdot \fd\tilde{\gamma_2}+\sin \tilde{\gamma_2} \cdot \sin (\tilde{\gamma_1}-\frac{\epsilon}{2})\cdot \fd \tilde{\gamma_3}\} \\
\fLambda_1&=-\frac{1}{2}\,\{\fd(\tilde{\gamma_1}-\frac{\epsilon}{2})+\cos \tilde{\gamma_2}\cdot \fd\tilde{\gamma_3}\}\\
\fX_2&=\frac{1}{2}\,\{\cos (\tilde{\gamma_1}+\frac{\epsilon}{2})\cdot \fd\tilde{\gamma_2}+\sin \tilde{\gamma_2} \cdot \sin (\tilde{\gamma_1}+\frac{\epsilon}{2})\cdot \fd \tilde{\gamma_3}\} \\
\fGamma_2&=\frac{1}{2}\,\{\sin \tilde{\gamma_2}\cdot \cos( \tilde{\gamma_1}+\frac{\epsilon}{2})\cdot \fd \tilde{\gamma}_3- \sin (\tilde{\gamma_1}+\frac{\epsilon}{2}) \cdot \fd \tilde{\gamma_2}\}\\
\fLambda_2&=\frac{1}{2}\,\{\fd(\tilde{\gamma_1}+\frac{\epsilon}{2})+\cos \tilde{\gamma_2}\cdot \fd\tilde{\gamma_3}\} \; .
\end{align}
\end{subequations}

On the other hand, these single-particle exchange fields may also be written in terms of the single-particle angles $\gamma_{j(a)}$ (\ref{e4:21}) via the Maurer-Cartan identifications (\ref{e4:17}) as ($\sigma_{\ast}=+1$)
\begin{subequations}
\label{e4:27}
\begin{align}
\fX_a & \equiv-\frac{1}{2}\,\fE^1{}_{(a)} =-\frac{1}{2}\,\{\sin \gamma_{1(a)} \cdot \fd \gamma_{2(a)}-\cos \gamma_{1(a)}\cdot \sin \gamma_{2(a)}\cdot \fd \gamma_{3(a)}\}\\
\fGamma_a & \equiv-\frac{1}{2}\,\fE^2{}_{(a)} = \frac{1}{2}\,\{\sin \gamma_{1(a)} \cdot \sin \gamma_{2(a)}\cdot \fd \gamma_{3(a)} + \cos \gamma_{1(a)}\cdot \fd\gamma_{2(a)}\}\\
\fLambda_a &\equiv\frac{1}{2}\, \fE^3{}_{(a)} =-\frac{1}{2}\,\{\fd\gamma_{1(a)}+\cos \gamma_{1(a)}\cdot \fd\gamma_{3(a)}\}\; .
\end{align}
\end{subequations}
Thus, comparing both versions (\ref{e4:26}) and (\ref{e4:27}) immediatly yields the desired exchange coupling conditions upon the Euler angles $\gamma_{j(a)}$ ($\sigma_{\ast}=+1$) in a very simple form:
\begin{subequations}
\label{e4:28}
\begin{align}
\gamma_{1(1)}&=\tilde{\gamma_1}-\frac{\epsilon}{2}\\
\gamma_{1(2)}&=\frac{\pi}{2}-(\tilde{\gamma_1}+\frac{\epsilon}{2})\\
\gamma_{2(1)}&=\tilde{\gamma_2}\\
\gamma_{2(2)}&=\pi-\tilde{\gamma_2}\\
\gamma_{3(1)}&=\gamma_{3(2)}=\tilde{\gamma_3} \; .
\end{align}
\end{subequations}
Clearly this coupling of the Euler angles $\gamma_{j(a)}$ actually represents a coupling of both group elements $\G_{a(x)}$ which necessarily must result also in the corresponding coupling of the Maurer-Cartan form $\fE^j{}_{(a)}$
\begin{subequations}
\label{e4:28_2}
\begin{align}
\fE^1{}_{(2)}&=\cos \epsilon \cdot \fE^2{}_{(1)}+\sin \epsilon \cdot \fE^1{}_{(1)}\\
\fE^2{}_{(2)}&=\cos \epsilon \cdot \fE^1{}_{(1)}-\sin \epsilon \cdot \fE^2{}_{(1)}\\
\fE^3{}_{(2)}&=\fE^3{}_{(1)}+\fd \epsilon \; .
\end{align}
\end{subequations}

Summarizing the present results with the Euler angles one encounters a very pleasant result; namely the parametrization of the exchange fields $\fX_a$, $\fGamma_a$, $\fLambda_a$ yields a general form of these exchange fields which automatically obeys those relatively complicated exchange coupling conditions (\ref{e3:20}); and additionally all the curl relations for the exchange fields are also obeyed automatically! Furthermore, the Euler angle parametrization provides one with a convenient possibility to incorporate certain symmetries into the desired RST solutions. For instance, imagine that we want to look for static solutions where all RST fields become time-independent, e.g. think of the bound solutions in an attractive potential. For such a situation one expects that all scalar fields, such as the Euler angles, cannot depend upon time $t$ but will exclusively depend on the space position ${\vec r}$. Moreover, the time-component of the four-vectors $X_{a \mu}$, $\Gamma_{a\mu}$, $\Lambda_{a\mu}$ must also be time-independent, e.g. we put
\begin{subequations}
\label{e4:34}
\begin{align}
X_{a \mu}&=X_a(\vec{r})\cdot \hat{t}_{\mu}\\ 
\Gamma_{a \mu}&=\Gamma_a(\vec{r})\cdot \hat{t}_{\mu} \; .
\end{align}
\end{subequations}
All these conditions can easily be satisfied by putting for the Euler angles 
\begin{subequations}
\label{e4:35}
\begin{align}
\epsilon&=\epsilon(\vec{r})\\
\tilde{\gamma_1}&=\tilde{\gamma_1}(\vec{r})\\
\tilde{\gamma_2}&=K_{\ast}\cdot t \quad \mbox{($K_{\ast}=$const.)}\\
\tilde{\gamma_3}&=\mbox{const.}
\end{align}
\end{subequations}
This then yields the static form of the exchange fields:
\begin{subequations}
\label{e4:36}
\begin{align}
X_{1 \mu}&=-\frac{1}{2}\; K_{\ast} \cosh (\tilde{\gamma_1}+\frac{\epsilon}{2}) \cdot \hat{t}_{\mu}\\
\Gamma_{1 \mu}&=-\frac{1}{2}\;K_{\ast} \sinh(\tilde{\gamma_1}+\frac{\epsilon}{2})\cdot \hat{t}_{\mu}\\
\Lambda_{1 \mu}&=-\frac{1}{2}\; \partial_{\mu} (\tilde{\gamma_1}+\frac{\epsilon}{2})\\
X_{2 \mu}&=-\frac{1}{2}\; K_{\ast} \cosh (\tilde{\gamma_1}-\frac{\epsilon}{2}) \cdot \hat{t}_{\mu}\\
\Gamma_{2 \mu}&=\frac{1}{2}\;K_{\ast} \sinh(\tilde{\gamma_1}-\frac{\epsilon}{2})\cdot \hat{t}_{\mu}\\
\Lambda_{2 \mu}&=\frac{1}{2}\; \partial_{\mu} (\tilde{\gamma_1}-\frac{\epsilon}{2})\; .
\end{align}
\end{subequations}
It was exactly this form of the exchange fields which has been applied for a treatment of the bound two-particle states in an attractive Coulomb force field \cite{RuSo01_2}.

\mysubsection{Three-Vector Parametrization}
For the subsequent computation of winding numbers it is very helpful to consider also an alternative parametrization of the group elements $\G_{(x)}$ (\ref{e4:20}), namely the parametrization by a ``three-vector'' $\{\xi^j;j=1,2,3\}$
\begin{equation}
\label{e4:37}
\G_{(x)}=\exp[\xi^j{}_{(x)}\tau_j] \; .
\end{equation}
It is true, expressing the exchange coupling condition in terms of this three-vector $\vec{\xi}$ is somewhat cumbersome; but on the other hand the parametrization (\ref{e4:37}) enables one to easily recognize the way how the topological quantum numbers come about in RST. Therefore it is instructive to consider the three-vector parametrization in some detail.

The adjoint transformation $G \rightarrow G$
\begin{equation}
\label{e4:38}
\G_{(x)}\rightarrow \G^{\prime}{}_{(x)}=\S\cdot \G_{(x)}\cdot \S^{-1}
\end{equation}
\centerline{($\S \in G$)}
acts over the groups $G=SO(3)$ and $G=SO(1,2)$ in a somewhat different way. Since the three-vector $\vec{\xi}=\{\xi^j\}$ changes under the transformation (\ref{e4:38}) according to
\begin{equation}
\label{e4:39}
\xi^j \rightarrow \xi^{\prime j}=S^j{}_k \xi^k \; ,
\end{equation}
where the matrix $\{ S^j{}_k\}$ is identical to $\S$, the three-vector $\vec{\xi}$ can be rotated into any direction in group space for $SO(3)$ but not for $SO(1,2)$. The reason is that the Killing metric $g_{jk}$ (\ref{e4:4}) is invariant with respect to these transformations i.e.
\begin{equation}
\label{e4:40}
\eta_{jk}S^j{}_lS^k{}_m=\eta_{lm} 
\end{equation}
and therefore the ``length'' of the three-vector is preserved
\begin{equation}
\label{e4:41}
\xi^{\prime j}\xi^{\prime}{}_j=\xi^j\xi_j \equiv \eta_{ij}\xi^i\xi^j \; .
\end{equation}
This condition does not restrict the possible end configurations of $\vec{\xi}$ for $SO(3)$ but it does for $SO(1,2)$. More concretely, since the Killing metric $\eta_{jk}$ (\ref{e4:16}) is indefinite for the negative mixtures ($\sigma_{\ast}=-1$), the three-vector $\vec{\xi}$ can never cross the ``light-cone'' $\xi^j\xi_j=0$ and thus the adjoint transformation (\ref{e4:38}) leaves invariant two three-dimensional subspaces in $SO(1,2)$, namely those which have $\xi^j\xi_j>0$ and $\xi^j\xi_j<0$. This is the reason why one has to resort to two different parametrizations of the exchange fields for the negative mixtures.

In detail, when $\vec{\xi}$ is ``time-like'' (i.e.  $\xi^j\xi_j>0$) the Maurer-Cartan components ${\Bbb E}^j_{(a)}$ (\ref{e4:11}) are found to be of the following form for the positive and negative mixtures ($\sigma_{\ast}=+1$)
\begin{equation}
\label{e4:42}
\fE^j=-u^j\fd\xi-\sin \xi \cdot \fd u^j+(\cos \xi -1)\epsilon^j{}_{kl}u^k\fd u^l \; .
\end{equation}
Here, a nearby decomposition of the three-vector $\vec{\xi}$ into a unit vector $\vec{u}$ ($u^j u_j=1$) and the ``length'' $\xi$ has been used, i.e.
\begin{equation}
\label{e4:43}
\xi^j=\xi \cdot u^j \; .
\end{equation}

\mysubsection{Change of Parametrization}
When different parametrizations are found to be advantageous for different purposes, it becomes necessary to specify the transformations which connect those various forms of parametrizations. The three-vector $\vec{\xi}$ parametrizing the group element $\G$ (\ref{e4:37}) can be expressed in terms of $\G$ itself in the following way: first extract the ``length'' $\xi$ of the three-vector through ($\sigma_{\ast}=+1$)
\begin{equation}
\label{e4:45}
\cos \xi =-1+\frac{1}{2} \tr \G
\end{equation}
and then find the unit vector $\vec{u}=\{u^j\}$ (\ref{e4:43}) through
\begin{equation}
\label{e4:46}
u^k\eta_{kj}=u_j=-\frac{1}{2\sin \xi} \tr (\G \cdot \tau_j) \; .
\end{equation}
Thus, referring to the Euler parametrization of $\G$ (\ref{e4:19})-(\ref{e4:20}), yields the length of the three-vector $\vec{\xi}$ in terms of the Euler angles as ($\sigma_{\ast}=+1$)
\begin{equation}
\label{e4:47}
\cos \xi=\frac{1}{2} \Big\{ \cos \gamma_2-1+(1+\cos \gamma_2)\cdot \cos(\gamma_1+\gamma_3) \Big\}
\end{equation}
and then the unit vector $\vec{u}=\{u^j\}$ is found from (\ref{e4:46}) as ($\sigma_{\ast}=+1$)
\begin{subequations}
\label{e4:48}
\begin{align}
u^1&=\frac{\sin \gamma_2}{2 \sin \xi} \Big( \sin \gamma_3 - \sin \gamma_1 \Big)\\
u^2&=\frac{\sin \gamma_2}{2 \sin \xi} \Big( \cos \gamma_1 + \cos \gamma_3 \Big)\\
u^3&=\frac{1+\cos \gamma_2}{2 \sin \xi} \sin (\gamma_1 +\gamma_3) \; .
\end{align}
\end{subequations}

The advantage of working with both parametrizations comes now into play when we transcribe the exchange coupling condition between both particles from the Euler parametrizations (\ref{e4:28}), where this condition looks very simple, to the three-vector parametrizations for which that coupling condition would have never been found directly. Thus, for the positive mixtures we obtain the three-vector $\vec{\xi}$ in terms of the average Euler angles $\tilde{\gamma_j}$ and exchange angle $\epsilon$ by combining the present result (\ref{e4:47})-(\ref{e4:48}) with the former coupling conditions (\ref{e4:28}) in order to find for the first particle ($\sigma_{\ast}=+1$)
\begin{subequations}
\label{e4:52}
\begin{align}
\cos \xi_{(1)}&=\frac{1}{2} \Big\{ \cos \tilde{\gamma_2} -1+(1+\cos \tilde{\gamma_2}) \cdot \cos (\tilde{\gamma_1}+\tilde{\gamma_3} -\frac{\epsilon}{2})\Big\}  \\
u^1{}_{(1)}&=\frac{\sin \tilde{\gamma_2}}{2 \sin \xi_{(1)}} \Big\{ \sin \tilde{\gamma_3} - \sin (\tilde{\gamma_1} -\frac{\epsilon}{2})\Big\} \\
u^2{}_{(1)}&=\frac{\sin \tilde{\gamma_2}}{2 \sin \xi_{(1)}} \Big\{ \cos (\tilde{\gamma_1}-\frac{\epsilon}{2})+\cos \tilde{\gamma_3} \Big\} \\
u^3{}_{(1)}&=\frac{1+\cos \tilde{\gamma_2}}{2 \sin \xi_{(1)}} \sin(\tilde{\gamma_1}+\tilde{\gamma_3}-\frac{\epsilon}{2})
\end{align}
\end{subequations}
and similarly for the second particle 
\begin{subequations}
\label{e4:53}
\begin{align}
\cos \xi_{(2)}&=\frac{1}{2} \Big\{ -(1+\cos \tilde{\gamma_2})+(1-\cos \tilde{\gamma_2}) \cdot \sin(\tilde{\gamma_3}-\tilde{\gamma_1}-\frac{\epsilon}{2}) \Big\}\\
u^1{}_{(2)}&=\frac{\sin \tilde{\gamma_2}}{2 \sin \xi_{(2)}} \Big\{ \sin \tilde{\gamma_3} -\cos(\tilde{\gamma_1} +\frac{\epsilon}{2}) \Big\}\\ 
u^2{}_{(2)}&=\frac{\sin \tilde{\gamma_2}}{2 \sin \xi_{(2)}} \Big\{ \cos \tilde{\gamma_3} + \sin (\tilde{\gamma_1}+\frac{\epsilon}{2}) \Big\}\\
u^3{}_{(2)}&=\frac{1-\cos \tilde{\gamma_2}}{2 \sin \xi_{(2)}} \cos (\tilde{\gamma_3} -\tilde{\gamma_1} -\frac{\epsilon}{2}) \; .
\end{align}
\end{subequations}

\newpage
\mysection{Winding Numbers ($\sigma_{\ast}=+1$)}
These results become now relevant for revealing the correlation of the topological quantum numbers of both particles where this correlation is a consequence of the exchange coupling. First recall that either particle requires its own map $\G_{a(x)}$ ($a=1,2$) from space-time to the exchange group $G$ (i.e. $SO(3)$ or $SO(1,2)$). However, on behalf of the exchange coupling conditions, both group elements $\G_{a(x)}$ become coupled as shown, e.g., by the present three-vector parametrizations (\ref{e4:52})-(\ref{e4:53}). This coupling of both group elements $\G_a$ can be viewed in such a way that an ``average'' group element $\tilde{\G}$ is introduced for both particles. This average $\tilde{\G}$ may be parametrized, e.g. by the average three-vector $\tilde{\xi}^j=\tilde{\xi}\cdot \tilde{u}^j$ and then the individual group elements $\G_a=\G_a(\xi^j{}_{(a)})$ are constructed from the average group element $\tilde{G} (\tilde{\xi}^j)$ by shifting the average $\tilde{\xi}^j$ to the individual $\xi^j{}_{(a)}$ for either particle as shown by the present results, e.g. (\ref{e4:52})-(\ref{e4:53}) for the positive mixtures. Clearly this latter map $\tilde{\G} \Rightarrow \G_a$ ($a=1,2$) is described by the left action of certain elements $\S_{(a)}$ upon the average group element $\tilde{\G}$. Thus, any three-cycle $C^3$ of space-time is first mapped into the exchange group $SO(3)$ via $\tilde{\G}(x)$ and thereby generates the ``average'' winding number $\tilde{Z}[C^3]$ as the number of times how often the exchange group is covered when the original $C^3$ is swept out once. Subsequently, the maps $\tilde{\G} \rightarrow \G_{(a)}$ mediated by the elements $\S_{(a)}$ generate the corresponding winding numbers $\tilde{l}_{(a)}$ in an analogous way, such that ultimately the exchange group becomes covered $\tilde{Z}\cdot \tilde{l}_{(a)}$ times, yielding the total winding numbers $Z_{(a)}$ ($a=1,2$) as 
\begin{equation}
\label{e5:1}
Z_{(a)}=\tilde{Z} \cdot \tilde{l}_{(a)} \; .
\end{equation}

On the other hand, one may consider the map from space-time to the exchange group $G=SO(3)$ with the first Euler angle $\tilde{\gamma_1}$ being kept constant, however the exchange angle $\epsilon$ being assumed as the given non-trivial space-time function. This map induces identical winding numbers ($Z_{\epsilon}$, say) for both particles, as may be seen from the exchange coupling (\ref{e4:28}). Since the average winding number $\tilde{Z}$ applies for the case when the exchange angle $\epsilon$ is kept constant, the general situation (\ref{e5:1}) is associated with the sum/difference of both winding numbers $\tilde{Z}$ and $Z_{\epsilon}$, i.e.
\begin{subequations}
\label{e5:2}
\begin{align}
Z_{(1)}&=\tilde{Z}\cdot \tilde{l}_{(1)} = \tilde{Z}-Z_{\epsilon}\\
Z_{(2)}&=\tilde{Z}\cdot \tilde{l}_{(2)} = \tilde{Z}+Z_{\epsilon} \; .
\end{align}
\end{subequations}
This however says for the ``interior'' winding numbers $\tilde{l}_{(a)}$
\begin{subequations}
\label{e5:3}
\begin{align}
\tilde{l}_{(1)} &=1-\frac{Z_{\epsilon}}{\tilde{Z}} \\
\tilde{l}_{(2)}&=1+\frac{Z_{\epsilon}}{\tilde{Z}} \; ,
\end{align}
\end{subequations}
i.e. the ratio of winding numbers $\frac{Z_{\epsilon}}{\tilde{Z}}$ must also be an integer ($\tilde{n}$, say)
\begin{equation}
\label{e5:4}
\frac{Z_{\epsilon}}{\tilde{Z}}\doteqdot \tilde{n} \; .
\end{equation}
Consequently the winding numbers $Z_{(a)}$ of both particles ($a=1,2$) are correlated in the following manner ($\tilde{n}=0,\pm 1,\pm 2,...$)
\begin{subequations}
\label{e5:6}
\begin{align}
Z_{(1)}&=(1-\tilde{n})\tilde{Z}\\
Z_{(2)}&=(1+\tilde{n})\tilde{Z} \; .
\end{align}
\end{subequations}
This result says that {\it both winding numbers $Z_{(a)}$ are either odd or even but are otherwise completely unrestricted}.

Subsequently we give a more rigorous treatment of these rather heuristic arguments.

\mysubsection{Invariant Volume}
It should be immediately obvious from the very definition of the Maurer-Cartan form $\tilde{\C}$ (\ref{e4:10}) that it transforms under the adjoint map when its generating group element $\G$ is acted upon by some (constant) element $\S$ of the exchange group $G$ 
\begin{subequations}
\label{e5:7}
\begin{align}
\G &\Rightarrow \G^{\prime}= \S \cdot \G\\
\tilde{\C}&\Rightarrow \tilde{\C}^{\prime}=\S\cdot\tilde{\C} \cdot \S^{-1} \; .
\end{align}
\end{subequations}
Therefore, referring to the adjoint representation of the exchange group $G$, it is possible to introduce a volume 3-form $\fV$ over $G$
\begin{equation}
\label{e5:8}
\fV=-g_{\ast} \tr (\C \wedge \C \wedge \C)
\end{equation}
with some real normalization constant $g_{\ast}$. Obviously, this volume form is invariant under the left action of $\S$ (\ref{e5:7}) and thus attributes an invariant volume $V$ to the exchange group $G$ if the latter is compact (i.e. for $SO(3)$ but not for $SO(1,2)$);
\begin{equation}
\label{e5:9}
V=\int_G \fV
\end{equation}
(Haar measure \cite{ChDeDi}).\newline
Decomposing here the pullback $\tilde{\C}$ of the Maurer-Cartan form $\C$ into its components $\fE^j$ (\ref{e4:11}) yields for the corresponding pullback $\tilde{\fV}$ of the volume form $\fV$
\begin{equation}
\label{e5:10}
\tilde{\fV}=-g_{\ast}\fE^j \wedge \fE^k \wedge \fE^l \cdot \tr(\tau_j\cdot \tau_k \cdot \tau_l)=g_{\ast}\epsilon_{ijk}\fE^i\wedge \fE^j \wedge \fE^k
\end{equation}
which applies to both kinds of mixtures ($\sigma_{\ast}=\pm 1$).

However, a crucial difference between both types of mixtures becomes now evident when expressing the volume form $\fV$ in terms of the Euler variables $\gamma_j$: namely, for the positive mixtures one finds by referring to their Maurer-Cartan forms (\ref{e4:21})
\begin{equation}
\label{e5:11}
\fV_+=3!\, g_{\ast} \sin \gamma_2 \fd\gamma_1 \wedge \fd \gamma_2 \wedge \fd\gamma_3
\end{equation}
\centerline{($\sigma_{\ast}=+1$)}
and similarly for the negative mixtures by use of their Maurer-Cartan forms (\ref{e4:22})
\begin{equation}
\label{e5:12}
\fV_-=3!\, g_{\ast} \sinh \gamma_2 \fd\gamma_1 \wedge \fd \gamma_2 \wedge \fd\gamma_3
\end{equation}
\centerline{($\sigma_{\ast}=-1$).}
Obviously the exchange group $SO(3)$ is covered once when the Euler angles sweep out their possible range of values $0<(\gamma_1$, $\gamma_3) \le 2 \pi$, $0 \le \gamma_2 \le \pi$, so that the invariant volume $V$ (\ref{e5:9}) becomes ($\sigma_{\ast}=1$)
\begin{equation}
\label{e5:13}
V=6g_{\ast}\int_0^\pi \sin \gamma_2 d\gamma_2 \int_0^{2\pi}d\gamma_1 \int_0^{2 \pi} d \gamma_3=48 g_{\ast} \pi^2 \; ,
\end{equation}
whereas such a definite volume $V$ cannot be attributed to the negative mixtures. 

Concerning the fixing of the normalization constant $g_{\ast}$, it is better to refer this to the universal covering group $SU(2)$ of $SO(3)$ because this has the same topology as the 3-cycle $C^3$ in space-time over which the pullback of the volume form $\fV$ has to be integrated over for any particle ($a=1,2$) in order to define its winding number $Z_{(a)}$ 
\begin{equation}
\label{e5:14}
Z_{(a)} = <\tilde{\fV}|C^3>\doteqdot \oint_{C^3}\tilde{\fV}
\end{equation}
as the number of times to cover $SO(3)$ when $C^3$ is swept out once. Since for a diffeomorphic map the universal covering group $SU(2)$ is swept out once when its homomorphic image $SO(3)$ is covered twice we put $g_{\ast}=(96\pi^2)^{-1}$ and thus find for the volume form $\fV$ (\ref{e5:11}) of the positive mixtures ($\sigma_{\ast}=+1$)
\begin{equation}
\label{e5:15}
\fV_{(a)}=\frac{1}{(4\pi)^2} \sin \gamma_{2(a)} \fd\gamma_{1(a)} \wedge \fd \gamma_{2(a)} \wedge \fd\gamma_{3(a)} \; .
\end{equation}

Once the volume forms $\fV$ of both particles have thus been obtained, it becomes easy to verify the preceding guess for the winding numbers $Z_{(a)}$ (\ref{e5:2})-(\ref{e5:3}). Indeed, one merely has to substitute the exchange coupling conditions (\ref{e4:28}) into those volume forms $\fV_{(a)}$ (\ref{e5:15}) and then to compute the individual winding numbers $Z_{(a)}$ (\ref{e5:14}). This yields immediately
\begin{subequations}
\label{e5:16}
\begin{align}
\tilde{Z}&=\frac{1}{(4\pi)^2} \oint_{C^3} \sin \tilde{\gamma_2} \fd \tilde{\gamma_1}\wedge \fd \tilde{\gamma_2} \wedge \fd \tilde{\gamma_3}\\
Z_{\epsilon}&=\frac{1}{(4\pi)^2}\oint_{C^3}\sin \tilde{\gamma_2} (\fd\frac{\epsilon}{2}) \wedge \fd \tilde{\gamma_2} \wedge \fd \tilde{\gamma_3} 
\end{align}
\end{subequations}
and thus the presumed result (\ref{e5:6}) is verified. Here the interior winding numbers $\tilde{l}_{(a)}$ (\ref{e5:1}) are given by 
\begin{equation}
\label{e5:17}
\tilde{l}_{(a)}=\oint_{SO(3)} \frac{\partial (\gamma_1,\gamma_2,\gamma_3)}{\partial (\tilde{\gamma_1},\tilde{\gamma_2},\tilde{\gamma_3})} \cdot \sin \tilde{\gamma_2} \fd \tilde{\gamma_1}\wedge \fd \tilde{\gamma_2} \wedge \fd \tilde{\gamma_3}\; ,
\end{equation}
where the Jacobian of the interior transformation $\tilde{\G} \rightarrow \G_a$ multiplies the invariant volume element of the exchange group $SO(3)$.

\mysubsection{Universal Covering group}
Properly speaking, the RST exchange fields do not refer to the exchange group itself but rather to its Lie algebra via the corresponding Maurer-Cartan form $\tilde{\C}$ (\ref{e4:10}). Therefore one can take as the exchange groups also the universal covering groups of $SO(3)$ and $SO(1,2)$, namely $SU(2)$ and $SU(1,1)$, resp. For the compact case, one constructs a unit four-vector $U^{\alpha}$ ($U^{\alpha}U_{\alpha}=\eta_{\alpha \beta}U^{\alpha}U^{\beta}=1$, $\eta_{\alpha \beta}=\mbox{diag} [1,1,1,1]$) from the three-vector $\vec{\xi}$ (\ref{e4:43}) in the following way:
\begin{subequations}
\label{e5:18}
\begin{align}
U^0&=\cos \frac{\xi}{2}\\
U^j&=\sin\frac{\xi}{2}\cdot u^j \;.
\end{align}
\end{subequations}
This yields the $SU(2)$ group element $\G(x)$ as 
\begin{equation}
\label{e5:19}
\G=U^0\cdot {\bf 1}-iU^j\sigma_j
\end{equation}
where $\sigma_j$ are the Pauli matrices as ususal. The associated Mauerer-Cartan form $\tilde{\C}$ (\ref{e4:10})-(\ref{e4:11}) reads now
\begin{equation}
\label{e5:20}
\C_{\mu}=-\frac{i}{2} E^j{}_{\mu} \sigma_j
\end{equation}
with the Maurer-Cartan forms $E^j{}_{\mu}$ given by (\ref{e4:42}). Clearly, for the negative mixtures one would introduce the unit four-vector $U^{\alpha}$ through
\begin{subequations}
\label{e5:21}
\begin{align}
U^0&=\cosh \frac{\xi}{2}\\
U^j&=\sinh \frac{\xi}{2}\cdot u^j 
\end{align}
\end{subequations} 
and would replace the $SU(2)$ generators $\tau_j(=-\frac{i}{2}\sigma_j)$ by the corresponding ${\mathfrak{su}}(1,1)$ generators in order to find the curresponding Maurer-Cartan forms.

The pleasant effect with the choice of the universal covering group $SU(2)$ for the positive mixtures is now that its nomalized volume form $\tilde{\fV}$ (\ref{e5:10})
\begin{equation}
\label{e5:22}
\tilde{\fV}=-\frac{1}{24 \pi^2} \tr (\tilde{\C} \wedge \tilde{\C} \wedge \tilde{\C}) =-\frac{1}{96\pi^2}\epsilon_{jkl}\fE^j \wedge \fE^k \wedge \fE^l 
\end{equation}
reads in the three-vector parametrization 
\begin{equation}
\label{e5:23}
\tilde{\fV}=\frac{1}{(2\pi)^2} \sin^2 \frac{\xi}{2} \fd \xi \wedge \Big\{\frac{1}{2} \epsilon_{jkl} u^j(\fd u^k)\wedge (\fd u^l) \Big\} \; ,
\end{equation}
or even more concisely in the four-vector parametrization

\begin{equation}
\label{e5:24}
\tilde{\fV}=\frac{3!}{2\pi^2} \epsilon_{\alpha \beta \gamma \delta} U^{\alpha}(\fd U^{\beta}) \wedge (\fd U^{\gamma}) \wedge (\fd U^{\delta}) \; .
\end{equation}
Obviously this is just the volume form over the 3-sphere $S^3$, parametrized by the unit four-vector $U^{\alpha}$ (\ref{e5:18}) and divided by the invariant volume $V=2 \pi^2$ (\ref{e5:9}) of the compact exchange group $SU(2)$. As a consequence we actually arrived at our original goal of obtaining integer winding numbers $Z_{(a)}$ (\ref{e5:14}) for any particle $a=1,2$, namely by specifying two group elements $\G_{(a)}$ of the $SU(2)$ form (\ref{e5:19}) parametrized by two four-vectors ($U_{(a)}{}^{\alpha}$) which however must be subjected to the exchange coupling condition. Thus we do now have at hand not only a group theoretical method of generating consistent pairs of exchange fields $\{ X_{a \mu}$, $\Gamma_{a \mu}$, $\Lambda_{a \mu}\}$ for our two-particle system but simultaneously we gained a method of classifying the corresponding two-particle solutions by an (even or odd) pair of topological quantum numbers.

\mysubsection{Kinematics of Exchange Coupling}
The transition to the universal covering groups does not only yield some further insight into the topology of the mixture configurations but it helps also to get a concrete geometric picture of the exchange coupling condition. This comes about by looking for the exchange constraints between  both four-vectors $U^{\alpha}{}_{(1)}$ and $U^{\alpha}{}_{(2)}$, as the group parameters for the exchange group elements $\G_{(1)}$ and $\G_{(2)}$ for either particle, cf. (\ref{e5:19}). In other words, we have to transcribe the former exchange coupling condition between the Euler angles (\ref{e4:28}) to the present four-vectors $U^{\alpha}{}_{(a)}$ (\ref{e5:18}), $a=1,2$. Clearly, this goal can be achieved by eliminating the average Euler angles $\tilde{\gamma_1}$, $\tilde{\gamma_2}$, $\tilde{\gamma_3}$ from the three-vector parametrizations (\ref{e4:52})-(\ref{e4:53}).

The point of departure is here again a convenient reparametrization of the four-vector components $U^{\alpha}{}_{(a)}$  so that the exchange coupling condition adopts a very simple shape when expressed in these new parameters. The key point is here a ($2+2$)-splitting of the group space, i.e. we put for either particle ($a=1,2$)
\begin{subequations}
\label{e5:25}
\begin{align}
(U_{||(a)})^2&=(U^1{}_{(a)})^2+(U^2{}_{(a)})^2\\
(U_{\perp (a)})^2&=(U^0{}_{(a)})^2+(U^3{}_{(a)})^2
\end{align}
\end{subequations} 
and thus consider the projections of the four-vectors $U^{\alpha}{}_{(a)}$ into the ($1,2$)-plane and the ($0,3$)-plane. Now for these projections one easily finds from the three-vector parametrization (\ref{e4:52})-(\ref{e4:53}) the following cross relations
\begin{subequations}
\label{e5:26}
\begin{align}
(U_{\perp (1)})^2&=(U_{||(2)})^2=\cos^2 \Big(\frac{\tilde{\gamma_2}}{2} \Big)\\  (U_{\perp (2)})^2&=(U_{||(1)})^2=\sin^2 \Big(\frac{\tilde{\gamma_2}}{2} \Big) \; .
\end{align}
\end{subequations} 
This means that when the projection of $U^{\alpha}{}_{(1)}$ into the ($0,3$)-plane is maximal, the corresponding projection of $U^{\alpha}{}_{(2)}$ is minimal and vice versa. Such an effect suggests to parametrize those projections in the following way (first particle, $a=1$)
\begin{subequations}
\label{e5:27}
\begin{align}
U^0{}_{(1)}&=\cos \Big(\frac{\tilde{\gamma_2}}{2} \Big)\cdot \cos \zeta_{(1)}\\
U^3{}_{(1)}&=\cos \Big(\frac{\tilde{\gamma_2}}{2} \Big)\cdot \sin \zeta_{(1)}\\
U^1{}_{(1)}&=\sin \Big(\frac{\tilde{\gamma_2}}{2} \Big)\cdot \cos \eta_{(1)}\\
U^2{}_{(1)}&=\sin \Big(\frac{\tilde{\gamma_2}}{2} \Big)\cdot \sin \eta_{(1)}\; ,
\end{align}
\end{subequations} 
and similarly for the second particle ($a=2$)
\begin{subequations}
\label{e5:28}
\begin{align}
U^0{}_{(2)}&=\sin \Big(\frac{\tilde{\gamma_2}}{2} \Big)\cdot \cos \zeta_{(2)}\\
U^3{}_{(2)}&=\sin \Big(\frac{\tilde{\gamma_2}}{2} \Big)\cdot \sin \zeta_{(2)}\\
U^1{}_{(2)}&=\cos \Big(\frac{\tilde{\gamma_2}}{2} \Big)\cdot \cos \eta_{(2)}\\
U^2{}_{(2)}&=\cos \Big(\frac{\tilde{\gamma_2}}{2} \Big)\cdot \sin \eta_{(2)}\; .
\end{align}
\end{subequations} 
Thus the remaining task is to reveal the exchange coupling condition upon the angles $\zeta_{(a)}$ and $\eta_{(a)}$ in the corresponding two-planes.

In the next step, the remaining two Euler angles $\tilde{\gamma_1}$ and $\tilde{\gamma_3}$ are expressed by the four-vector components $U^{\alpha}{}_{(a)}$ in the following way
\begin{subequations}
\label{e5:29}
\begin{align}
\cos \Big(\tilde{\gamma_1}+\tilde{\gamma_3}-\frac{\epsilon}{2}\Big)&=\frac{(U^0{}_{(1)})^2-(U^3{}_{(1)})^2}{(U^0{}_{(1)})^2+(U^3{}_{(1)})^2}=\cos (2 \zeta_{(1)} )\\
\sin \Big(\tilde{\gamma_1}+\tilde{\gamma_3}-\frac{\epsilon}{2}\Big)&=2\, \frac{U^0{}_{(1)}\cdot U^3{}_{(1)}}{(U_{\perp}{}_{(1)})^2} =\sin (2 \zeta_{(1)})\\
\sin \Big(\tilde{\gamma_3}-\tilde{\gamma_1}-\frac{\epsilon}{2}\Big)&=-\frac{(U^0{}_{(2)})^2-(U^3{}_{(2)})^2}{(U^0{}_{(2)})^2+(U^3{}_{(2)})^2}=-\cos (2 \zeta_{(2)} )\\
\cos \Big(\tilde{\gamma_3}-\tilde{\gamma_1}-\frac{\epsilon}{2}\Big)&=2\, \frac{U^0{}_{(2)}\cdot U^3{}_{(2)}}{(U_{\perp}{}_{(2)})^2} =\sin (2 \zeta_{(2)})\; .
\end{align}
\end{subequations} 
From here one recognizes immediately that the angles $\zeta_{(a)}$ in the ($0,3$)-plane are linked to the Euler angles $\tilde{\gamma_1}$, $\tilde{\gamma_3}$ as follows
\begin{subequations}
\label{e5:30}
\begin{align}
2 \zeta_{(1)}&=\tilde{\gamma_1}+\tilde{\gamma_3}-\frac{\epsilon}{2}\\
2 \zeta_{(2)}&=\frac{\pi}{2}-\tilde{\gamma_1}+\tilde{\gamma_3}-\frac{\epsilon}{2} \; . 
\end{align}
\end{subequations} 

Finally, the last step consists in eliminating one of the Euler angles $\tilde{\gamma_1}$ or $\tilde{\gamma_3}$ from the first and second components $U^1{}_{(a)}$, $U^2{}_{(a)}$ of both particles. Thus, eliminating $\tilde{\gamma_1}$ yields by use of (\ref{e5:30}) the following two equations
\begin{subequations}
\label{e5:31}
\begin{align}
\zeta_{(1)}&=\eta_{(1)}+\tilde{\gamma_3}-\frac{\pi}{2}\\
\zeta_{(2)}&=-\frac{\pi}{2}+\eta_{(2)}+\tilde{\gamma_3} \; , 
\end{align}
\end{subequations} 
and, similarly, eliminating $\tilde{\gamma_3}$ yields
\begin{subequations}
\label{e5:32}
\begin{align}
\zeta_{(1)}&=\tilde{\gamma_1}-\eta_{(1)}-\frac{\epsilon}{2}+\frac{\pi}{2} \\
\zeta_{(2)}&=-\tilde{\gamma_1}-\eta_{(2)}-\frac{\epsilon}{2}-\pi \; . 
\end{align}
\end{subequations}
Clearly, both sets of equations (\ref{e5:31}) and (\ref{e5:32}) must be consistent when the former relationships are to be respected, which provides us with the final form of exchange coupling:
\begin{subequations}
\label{e5:33}
\begin{align}
\zeta_{(2)}&=\frac{3\pi}{4}-\eta_{(1)}-\frac{\epsilon}{2}\\
\eta_{(2)}&=-\zeta_{(1)}-\frac{\epsilon}{2}-\frac{\pi}{4} \; . 
\end{align}
\end{subequations}

This pleasant result allows us now to explicitly construct pairs of exchange fields which obey {\it both} the required curl relations {\it and} the exchange coupling conditions! Obviously we simply can {\it choose} one of both four-vectors $U^{\alpha}{}_{(x)}$ over space time, say $U^{\alpha}{}_{(1)}$, and then the second four-vector $U^{\alpha}{}_{(2)}$ is immediately given by equations (\ref{e5:28}) together with (\ref{e5:33}). Clearly the passage from the chosen $U^{\alpha}{}_{(1)}$ to the second vector $U^{\alpha}{}_{(2)}$ is still parametrized by the exchange angle $\epsilon$ which however can also be chosen arbitrarily! This may be seen more explicitly by recasting the transition from the first particle (\ref{e5:27}) to the second particle (\ref{e5:28}) into the form of an $SO(4)$ rotation
\begin{equation}
\label{e5:34}
U^{\alpha}{}_{(2)}=S^{\alpha}{}_{\beta}\,U^{\beta}{}_{(1)} \; .
\end{equation}
Here the rotation matrix $S^{\alpha}{}_{\beta}(\epsilon)$ is given by two $O(2)$ elements $\R$ through
\begin{equation}
\label{e5:35}
S^{\alpha}{}_{\beta}=\left( \begin{array}{cc} 0 &\R\big(\frac{\epsilon}{2}+\frac{\pi}{4}\big) \\ \R \big(\frac{\epsilon}{2}-\frac{3\pi}{4}\big) &0 \end{array} \right)
\end{equation}
with the $O(2)$ element $\R$ being given by
\begin{equation}
\label{e5:36}
\R_{(\alpha)}=\left( \begin{array}{cc} \cos \alpha & -\sin \alpha \\ -\sin \alpha & -\cos \alpha \end{array} \right) \; .
\end{equation}

\newpage
\mysection{Example: Stereographic Projection}
As the simplest demonstration of the present results, one may put to zero the exchange angle $\epsilon$ which then turns the $SO(4)$ matrix $\S$ (\ref{e5:34})-(\ref{e5:36}) into a constant matrix $\oS$ 
\begin{equation}
\label{e6:1}
\oS=\left( \begin{array}{cc} 0 & \R_{\left(\frac{\pi}{4}\right)} \\ \R_{\left(-\frac{3 \pi}{4}\right)} & 0 \end{array} \right) \; .
\end{equation}
For such a situation, both winding numbers $Z_{(a)}$ (\ref{e5:14}) must necessarily coincide with the average winding number $\tilde{Z}$ because the winding number $Z_{\epsilon}$ (\ref{e5:16}b) is trivially zero. The coincidence of $Z_{(a)}$ and $\tilde{Z}$ may also be seen from the four-vector version (\ref{e5:24}) of the volume three-forms $\fV_{(a)}$ since the constant $SO(4)$ element $\oS$ escapes the exterior differentiation and just reproduces the $SO(4)$ invariant permutation tensor
\begin{equation}
\label{e6:2}
\epsilon_{\mu \nu \lambda \sigma}= (\det \oS) \cdot \oS^{\alpha}{}_{\mu}\cdot \oS^{\beta}{}_{\nu} \cdot \oS^{\gamma}{}_{\lambda}\cdot \oS^{\delta}{}_{\sigma} \cdot \epsilon_{\alpha \beta \gamma \delta} \; .
\end{equation}
Clearly $\det \oS=+1$ because $\oS$ is an element of the proper rotation group.

\mysubsection{Stereographic Projection}
Thus we are left with the problem of determining some differentiable four-vector field $U^{\alpha}{}_{(1)}$ over space-time from which the second four-vector $U^{\alpha}{}_{(2)}$ can then be simply constructed by means of the $SO(4)$ rotation $\oS$ (\ref{e6:1}), cf. (\ref{e5:34}). Here, it is well-known that Euclidean 3-space (as a time slice of space-time $x^0=\mbox{const.}$) can be compactified to the 3-sphere $S^3$ by stereographic projection \cite{GoSc}. This procedure yields the following differentiable four-vector field $U^{\alpha}{}_{(x)}$ over $E_3$:
\begin{subequations}
\label{e6:3}
\begin{align}
U^0{}_{(1)}&=\frac{r^2-a^2}{r^2+a^2}\\
U^j{}_{(1)}&=2a^2\frac{x^j}{a^2+r^2} \; .
\end{align}
\end{subequations} 
This immediately yields for the average Euler angle $\tilde{\gamma_2}$ (\ref{e5:27})
\begin{equation}
\label{e6:4}
\sin \big( \frac{\tilde{\gamma_2}}{2}\big)=\frac{2 a r \sin \theta}{a^2+r^2}
\end{equation}
where spherical polar coordinates ($r$, $\theta$, $\phi$) have been used in place of the Cartesian parameters $\{x^j\}$. Hence for spatial infinity ($r=\infty$) the four-vector $U^{\alpha}{}_{(1)}$ points to the north pole of $S^3$ ($\tilde{\gamma_2}=0$), and for the origin ($r=0$) of $E_3$ to the south pole ($\tilde{\gamma_2}=\pi$), see fig. 2.

Once the first four-vector $U^{\alpha}{}_{(1)}$ has been fixed now, the corresponding second four-vector $U^{\alpha}{}_{(2)}$ can be constructed in a straight-forward manner by means of the $SO(4)$ rotation process
\begin{subequations}
\label{e6:5}
\begin{align}
U^0{}_{(2)}&=\frac{2ar \sin \theta}{a^2+r^2} \cos \big(\phi-\frac{3 \pi}{4} \big)\\
U^3{}_{(2)}&=-\frac{2ar \sin \theta}{a^2+r^2} \sin \big(\phi-\frac{3 \pi}{4} \big)\\
U^1{}_{(2)}&=\frac{1}{\sqrt{2}}\cdot \frac{r^2-a^2-2ar \cos \theta}{a^2+r^2}\\
U^2{}_{(2)}&=-\frac{1}{\sqrt{2}}\cdot \frac{r^2-a^2+2ar \cos \theta}{a^2+r^2}\; .
\end{align}
\end{subequations} 
Observe here that the second four-vector $U^{\alpha}{}_{(2)}$ (\ref{e6:5}) cannot be made identical to the first one $U^{\alpha}{}_{(1)}$ (\ref{e6:3}), because the exchange coupling condition always requires a {\it non-trivial} $SO(4)$ rotation $\S$ (\ref{e5:35})!

\mysubsection{Winding Numbers}
Since for the present exchange configuration both winding numbers coincide with the average winding number $\tilde{Z}$($=Z_{(1)}=Z_{(2))}$), it may be sufficient here to consider only the first number  $Z_{(1)}$. The corresponding volume three-form $\tilde{\fV}_{(1)}$ (\ref{e5:23}) reads
\begin{equation}
\label{e6:6}
\tilde{\fV}_{(1)}=\frac{1}{(2\pi)^2} \sin^2 \frac{\xi_{(1)}}{2}\, \fd\xi_{(1)} \wedge \big\{ \frac{1}{2} \epsilon_{j k l}u^j{}_{(1)}\, \fd u^k{}_{(1)} \wedge \fd u^l{}_{(1)} \big\} 
\end{equation}
where the three-vector length $\xi_{(1)}$ is obtained from the four-vector $U^{\alpha}{}_{(1)}$ (\ref{e5:18}) through
\begin{equation}
\label{e6:7}
\sin^2 \frac{\xi_{(1)}}{2}=U^j{}_{(1)}U_{j(1)}=1-(U^0{}_{(1)})^2 \; . 
\end{equation}
Thus, when the four-vector $U^{\alpha}{}_{(1)}$ is generated by stereographic projection as shown by equations (\ref{e6:3}), one finds for $\xi_{(1)}$
\begin{subequations}
\label{e6:8}
\begin{align}
\sin \frac{\xi_{(1)}}{2}=\frac{2ar}{a^2+r^2} \\
\cos \frac{\xi_{(1)}}{2}=\frac{a^2-r^2}{a^2+r^2}\; , 
\end{align}
\end{subequations}
i.e. we put $\xi_{(1)}=2 \pi$ for $r=0$ and $\xi_{(1)}=0$ for $r=\infty$. Furthermore, the unit three-vector $u^j{}_{(1)}$ simply becomes the normalized position vector of the Euclidean three-plane ($x^0=\mbox{const.}$)
\begin{equation}
\label{e6:9}
u^j{}_{(1)}=\frac{x^j}{r} \doteqdot \hat{x}^j \; .
\end{equation}

Consequently, the general volume form: $\tilde{\fV}_{(1)}$ (\ref{e6:6}) becomes specialized here to 
\begin{equation}
\label{e6:10}
\tilde{\fV}_{(1)}=-\Big(\frac{2}{\pi}\Big)^2 \frac{a^3}{(a^2+r^2)^3}\,r^2\fd r \wedge (\sin \theta\, \fd \theta\wedge \fd \phi)
\end{equation}
and obviously contains as its angular part the volume two-form over the sphere $S^2$. The integrations can therefore be done in a straightforward way and yield the expected result for the winding number $Z_{(1)}$ ($=Z_{(2)}$)
\begin{equation}
\label{e6:11}
Z_{(1)}=-\Big(\frac{2}{\pi}\Big)^2 a^3 \int_{r=o}^{\infty}\frac{r^2 dr}{(a^2+r^2)^3} \int_{\theta=0}^{\pi} d\theta \sin \theta \int_{\phi=0}^{2 \pi} d\phi=-1 \; .
\end{equation}
Clearly the Euclidean 3-space $E_3$ is wound up to the 3-sphere $S^3$ just once by the stereographic projection, whereby spatial infinity ($r=\infty$) is mapped to the north pole (which reverses the conventional orientation and gives a minus sign for the winding number).

\mysubsection{Exchange Fields}
Once the generating group elements $\G_a$ of the $SU(2)$ form (\ref{e5:19}) have been determined with the corresponding four-vector parametrizations $U^{\alpha}{}_{(a)}$ being displayed through equations (\ref{e6:3}) and (\ref{e6:5}), it is an easy matter to compute the associated Maurer-Cartan forms $\fE^j{}_{(a)}$. Here it is convenient to resort to the three-vector parametrization (\ref{e4:42}) of the Maurer-Cartan forms where the length $\xi_{(1)}$ of the first three-vector $\vec{\xi}_{(1)}$ (\ref{e4:43}) has been specified through equations (\ref{e6:8}) and the unit three-vector $u^j{}_{(1)}$ is given by (\ref{e6:9}). With these prerequisites the Maurer-Cartan forms for the first particle are found to be of the following form
\begin{equation}
\label{e6:12}
\fE^j{}_{(1)}=\frac{4a}{a^2+r^2} \hat{x}^j \fd r-4ar\frac{a^2-r^2}{(a^2+r^2)^2}\fd\hat{x}^j-8\frac{a^2r^2}{(a^2+r^2)^2}\epsilon^j{}_{kl}\hat{x}^k\fd\hat{x}^l \; .
\end{equation}

Concerning the Maurer-Cartan forms of the second particle $\fE^j{}_{(2)}$, one can refer to their link (\ref{e4:28_2}) to the first particle's expressions $\fE^j{}_{(1)}$ with the present specialization of putting the exchange angle to zero. This immediately yields for the second particle 
\begin{subequations}
\label{e6:13}
\begin{align}
\fE^1{}_{(2)}&=\fE^2{}_{(1)}\\
\fE^2{}_{(2)}&=\fE^1{}_{(1)}\\
\fE^3{}_{(2)}&=-\fE^3{}_{(1)} \; ,
\end{align}
\end{subequations} 
or explicitly
\begin{subequations}
\label{e6:14}
\begin{align}
\fE^1{}_{(2)}&\equiv -2 \fX_2 = \frac{4a}{a^2+r^2}\hat{x}^2\fd r-4ar\frac{a^2-r^2}{(a^2+r^2)^2}\fd\hat{x}^2-8\frac{a^2r^2}{(a^2+r^2)^2}[\hat{x}^3\fd\hat{x}^1-\hat{x}^1\fd\hat{x}^3]\\
\fE^2{}_{(2)}&\equiv -2 \fGamma_2 = \frac{4a}{a^2+r^2}\hat{x}^1\fd r-4ar\frac{a^2-r^2}{(a^2+r^2)^2}\fd\hat{x}^1-8\frac{a^2r^2}{(a^2+r^2)^2}[\hat{x}^2\fd\hat{x}^3-\hat{x}^3\fd\hat{x}^2]\\
\fE^3{}_{(2)}&\equiv 2 \fLambda_2 = -\frac{4a}{a^2+r^2}\hat{x}^3\fd r+4ar\frac{a^2-r^2}{(a^2+r^2)^2}\fd\hat{x}^3+8\frac{a^2r^2}{(a^2+r^2)^2}[\hat{x}^1\fd\hat{x}^2-\hat{x}^2\fd\hat{x}^1] \; .
\end{align}
\end{subequations}
Observe here that for the third form $\fE^3{}_{(a)}\equiv 2\fLambda_{(a)}$ (\ref{e6:13}c) we have on account of the constancy of the exchange angle $\epsilon$ the relationship
\begin{equation}
\label{e6:15}
\fLambda_1=-\fLambda_2
\end{equation}
which of course just meets with the former derivative of $\epsilon$ (\ref{e3:39}b) for vanishing $\epsilon$. Furthermore one can show that for vanishing $\epsilon$ the mixture variable $\beta$ must also vanish and therefore the exchange coupling (\ref{e3:20}) says:
\begin{subequations}
\label{e6:16}
\begin{align}
X_{1 \mu}&\equiv \Gamma_{2 \mu}\\
X_{2 \mu}&\equiv \Gamma_{1 \mu}
\end{align}
\end{subequations}
which is nothing else than the above result (\ref{e6:13}).

\mysubsection{Single-Particle Fields}
In Sect. III it has been made clear that the total set of RST variables can be subdivided into the subset of single-particle fields $\{{\Bbb L}_a$, $\KR_{a \mu} \}$ and exchange fields $\{X_{a \mu}$, $\Gamma_{a \mu}$, $\Lambda_{a \mu}\}$ such that both sets become coupled through the specific nature of the RST dynamics. Now that the exchange fields have been determined by hand in order to clearly display their topological features, one can face the problem of determining the dynamically associated single-particle fields $\{ {\Bbb L}_a$, $\KR_{a \mu}\}$ by solving the single-particle dynamics on the background of the given exchange fields. However, since the RST field equations constitute a very complicated dynamical system, one cannot hope to attain here an exact solution where the single-particle fields would be of a comparably simple form as the present exchange fields (\ref{e6:12}) or (\ref{e6:14}). Rather, we will be satisfied with convincing ourselves of the consistency of the RST dynamics so that one can be sure that the desired single-particle solution, to be associated to our construction of the exchange field system, does exist in principle.

First, consider the dynamical equations for the kinetic fields $\KR_{a \mu}$ which must consist of a source equation and of a curl equation in order to fix these vector fields, apart from certain boundary conditions. The source equations consist of both equations (\ref{e3:36}a), where however the coupling conditions (\ref{e6:15})-(\ref{e6:16}) are to be taken into account:
\begin{subequations}
\label{e6:17}
\begin{align}
\nabla^{\mu}\KR_{1 \mu}+2 {\Bbb L}_1{}^{\mu}\KR_{1 \mu}&=-2 X_1{}^{\mu}({\Bbb L}_{1 \mu}-{\Bbb L}_{2 \mu})-2 \Lambda_1{}^{\mu}\KR_{2 \mu} \\
\nabla^{\mu}\KR_{2 \mu}+2 {\Bbb L}_2{}^{\mu}\KR_{2 \mu}&=2\Gamma_1{}^{\mu}({\Bbb L}_{1 \mu}-{\Bbb L}_{2 \mu})+2 \Lambda_1{}^{\mu}\KR_{1 \mu} \; .
\end{align}
\end{subequations}
Furthermore, the curl equations for the kinetic fields have already been specified by equations (\ref{e3:35a}) where the gauge fields $F_{a \mu \nu}$ emerging there couple back to the RST currents $j_{a \mu}$ (\ref{e3:28}) via the Maxwell equations (\ref{e2:16}), i.e. for the present two-particle situation
\begin{equation}
\label{e6:18}
\nabla^{\mu}F_{a \mu \nu}=4 \pi \alpha_{\ast}j_{a \nu} \; .
\end{equation}
Observe here that the RST currents $j_{a \mu}$ themselves are to be constructed by means of the single-particle and exchange fields according to 
\begin{equation}
\label{e6:19}
j_{a \mu}=\frac{\hbar}{M c}{\Bbb L}_a{}^2 \{ \KR_{a \mu}+X_{a \mu}\}  \; .
\end{equation}

Evidently even if the exchange fields $X_{a \mu}$, $\Gamma_{a \mu}$, $\Lambda_{a \mu}$ are known (e.g. by our procedure of stereographic projection (\ref{e6:12})-(\ref{e6:14})), the kinetic fields still couple through their sources (\ref{e6:17}) to the unknown amplitude fields ${\Bbb L}_a$. Therefore the single-particle dynamics must be closed by specifying the field equations for the amplitudes ${\Bbb L}_a$ which are of course given by the wave equations (\ref{e3:32}) where the known exchange fields have to be inserted:
\begin{equation}
\label{e6:20}
\Box {\Bbb L}_a +{\Bbb L}_a \Big\{ \left(\frac{Mc}{\hbar}\right)^2 -\KR_{a \mu}\KR_a{}^{\mu} - 2 \KR_{a \mu}X_a{}^{\mu}\Big\}=\sigma_{\ast}{\Bbb L}_a \{ \Gamma_{a \mu}\Gamma_a{}^{\mu}+\Lambda_{a \mu}\Lambda_a{}^{\mu}+\sigma_{\ast} \X_{a \mu}X_a{}^{\mu}\} \; . 
\end{equation}
The right-hand side may be interpreted as the effect of an ``exchange potential'' $Y_a$
\begin{equation}
\label{e6:21}
Y_a \doteqdot \Gamma_{a \mu}\Gamma_a{}^{\mu}+\Lambda_{a \mu}\Lambda_a{}^{\mu}+\sigma_{\ast} \X_{a \mu}X_a{}^{\mu} \; . 
\end{equation}
where however both potentials are identical ($Y_1\equiv Y_2$) because of the general quadratic exchange coupling (\ref{e3:40}). There is no doubt that the present single-particle dynamics will admit non-trivial solutions which are however too difficult to be constructed analytically (for constructing {\it approximate} static solutions see ref. \cite{RuSo01_2}).

\newpage

\vspace{0.5cm}

\begin{figure}
\begin{center}
\epsfig{file=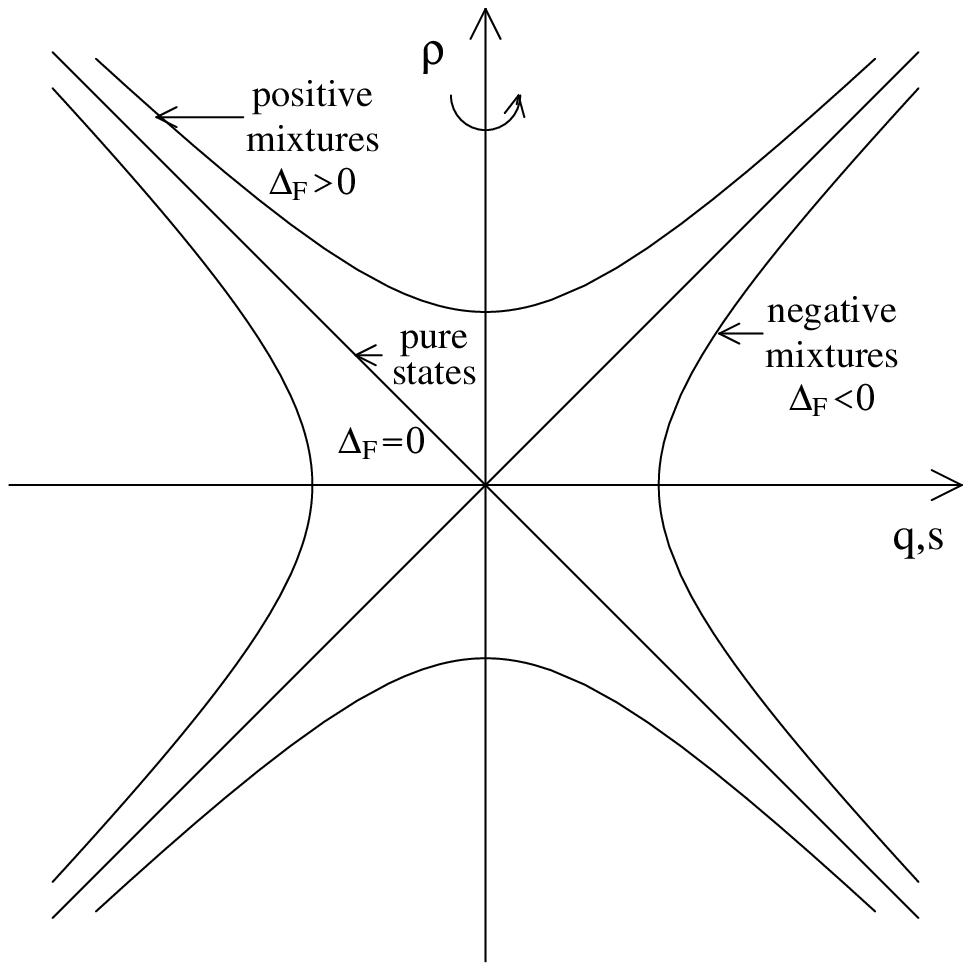}
\end{center}
\caption{The relativistic von Neumann equation (\ref{e2:2}) subdivides the density configuration
space into three subsets: the {\em pure states} occupy the {\em Fierz
  cone} $(\Delta_{F}=0)$, {\em positive mixtures} $(\Delta_{F}>0)$ are geometrically
represented by the two-parted hyperboloid and the {\em negative mixtures}
$(\Delta_{F}<0)$ by the one-parted hyperboloid. The mixtures approach the
pure states for $\Delta_{F}\rightarrow 0$. The general RST dynamics forbids a
change of the mixture type, cf. (\ref{e3:10}). The positive (negative)
mixtures may be considered as the RST counterparts of the symmetric and
anti-symmetric states of the conventional quantum theory.}
\label{fig2}
\end{figure}

\newpage

\begin{figure}
\begin{center}
\epsfig{file=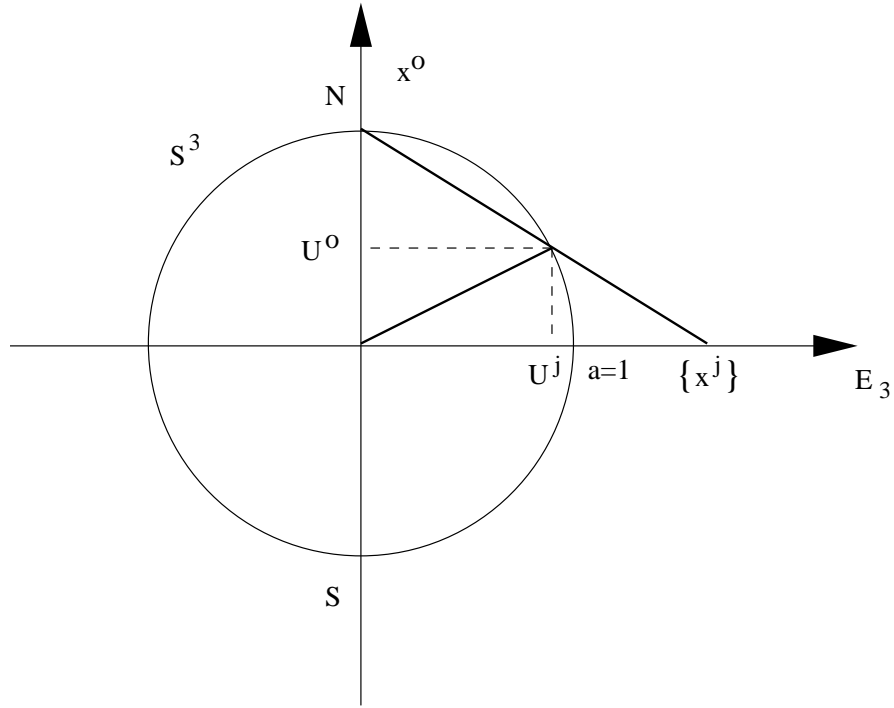}
\end{center}
\caption{{\it Stereographic projection}. Any point $\{x^j\}$ of Euclidean 3-space $E_3$ is mapped onto the three sphere $S^3$ $(x^0)^2+x^jx_j=1$ by projection from the north pole N $\{x^0=a \mbox{; }x^j=0\}$. This compactifies $E_3$ by identifying three-infinity ($r=\infty$) with the north pole N of $S^3$, whereas the origin ($r=0$) of $E_3$ is mapped into the south pole S.}
\label{fig3}
\end{figure}

\end{document}